\documentclass[acmsmall, screen]{acmart}

\usepackage{multirow}
\usepackage{comment}
\usepackage{listings}
\usepackage{xcolor}
\usepackage{balance}
\usepackage{tcolorbox}
\usepackage{listings}
\usepackage{tablefootnote}
\usepackage[normalem]{ulem}\usepackage{listings}
\usepackage{xcolor}
\usepackage{colortbl}
\definecolor{codegreen}{rgb}{0,0.6,0}
\definecolor{codegray}{rgb}{0.5,0.5,0.5}
\definecolor{codepurple}{rgb}{0.58,0,0.82}
\definecolor{backcolour}{rgb}{0.95,0.95,0.92}
\usepackage{longtable}
\usepackage{graphicx}
\usepackage{subcaption}
\lstdefinestyle{mystyle}{
  label=code:sample,
  frame=single,
  backgroundcolor=\color{backcolour}, commentstyle=\color{codegreen},
  keywordstyle=\color{magenta},
  numberstyle=\tiny\color{codegray},
  stringstyle=\color{codepurple},
  basicstyle=\ttfamily\footnotesize,
  breakatwhitespace=false,         
  breaklines=true,                 
  captionpos=t,                    
  keepspaces=true,                 
  numbers=left,                    
  numbersep=5pt,                  
  showspaces=false,                
  showstringspaces=false,
  showtabs=false,                  
  tabsize=1
}

\definecolor{mGreen}{rgb}{0,0.9,0}
\definecolor{mGray}{rgb}{0.5,0.5,0.5}
\definecolor{lighGray}{rgb}{0.8,0.8,0.8}
\definecolor{mPurple}{rgb}{0.58,0,0.82}
\definecolor{backgroundColour}{rgb}{0.95,0.95,0.92}
\usepackage{soul}
\lstdefinestyle{CStyle}{
    backgroundcolor=\color{backgroundColour},   
    commentstyle=\color{mGreen},
    keywordstyle=\color{magenta},
    numberstyle=\tiny\color{mGray},
    stringstyle=\color{mPurple},
    basicstyle=\footnotesize,
    breakatwhitespace=false,         
    breaklines=true,                 
    captionpos=b,                    
    keepspaces=true,                 
    numbers=left,                    
    numbersep=5pt,                  
    showspaces=false,                
    showstringspaces=false,
    showtabs=false,                  
    tabsize=2,
    language=C
}
\usepackage{array}
\usepackage{enumitem}
\usepackage{multirow}
\usepackage[ruled,linesnumbered]{algorithm2e}
\usepackage{bm}
\usepackage{url}
\usepackage[colorinlistoftodos]{todonotes}
\SetAlFnt{\small}
\SetAlCapFnt{\small}
\SetAlCapNameFnt{\small}
\usepackage{algorithmic}
\algsetup{linenosize=\small}
\SetKwInput{KwInput}{Input}                
\SetKwInput{KwOutput}{Output}              

\newcommand{\find}[1]{
\begin{tcolorbox}[leftrule=1mm,toprule=0mm,bottomrule=0mm,left=1pt,right=2pt,top=2pt,bottom=2pt]
\em #1
\end{tcolorbox}
}

\newboolean{showcomments}
\setboolean{showcomments}{true}
\ifthenelse{\boolean{showcomments}}
{ }

\ifthenelse{\boolean{showcomments}}
{ }

\ifthenelse{\boolean{showcomments}}
{ }

\usepackage{caption}
\newcommand{\thanh}[1]{#1}

\AtBeginDocument{%
  }

\setcopyright{acmcopyright}
\copyrightyear{2023}
\acmYear{2023}
\acmDOI{XXXXXXX.XXXXXXX}

\begin{document}

\title{Towards Reliable Evaluation of Neural Program Repair  with Natural Robustness Testing}


\author{Thanh Le-Cong}
\email{congthanh.le@student.unimelb.edu.au}
\affiliation{%
  \institution{The University of Melbourne}
  \country{Australia}
}

\author{Dat Nguyen}
\email{thanhdatn@student.unimelb.edu.au}
\affiliation{%
  \institution{The University of Melbourne}
  \country{Australia}
}

\author{Bach Le}
\email{bach.le@unimelb.edu.au}
\affiliation{%
  \institution{The University of Melbourne}
  \country{Australia}
}

\author{Toby Murray}
\email{toby.murray@unimelb.edu.au}
\affiliation{%
  \institution{The University of Melbourne}
  \country{Australia}
}

\renewcommand{\shortauthors}{Le-Cong et al.}

\begin{abstract}
Automated program repair (APR) has recently been gaining ground with numerous research efforts spent in the area that have been adopted in the industry. One notable class of APR is neural program repair (NPR), which typically employs deep learning techniques that are trained on vast amounts of historical data to fix bugs unseen in the past. 
To study the true effectiveness of NPR on existing limited dataset, recent works augment the evaluation data by employing semantics-preserving transformations to convert original buggy programs to semantically equivalent ones. Experiments show that NPR techniques are not robust, e.g., NPR cannot repair semantically-equivalent counterparts of 20\%-35\% of bugs that they can repair in the original dataset. \thanh{However, we found that many of these transformations are unnatural, that are unlikely to occur in real-world scenarios, leading to misleading conclusions about NPR effectiveness and misguide the improvement on unrobust behaviors, which have minimal real-world impact.}

\thanh{In this paper, we propose shifting the focus of robustness evaluation for Neural Program Repair (NPR) techniques towards naturally-occurring data transformations.} To accomplish this, we first examine the naturalness of semantic-preserving transformations through a two-stage human study. This study includes: (1) interviews with senior software developers to establish concrete criteria for evaluating the naturalness of these transformations, and (2) a survey involving 10 developers to assess the naturalness of 1,178 transformations, i.e., pairs of original and transformed programs, applied to 225 real-world bugs. Our findings show that only 60\% of these transformations are deemed natural, while 20\% are considered unnatural, with strong agreement among annotators. \thanh{Moreover, the unnaturalness of these transformations significantly impacts both their applicability to benchmarks and the conclusions drawn from robustness testing.} Next, we conduct natural robustness testing on NPR techniques to assess their true effectiveness against real-world data variations. Our experimental results reveal a substantial number of prediction changes in NPR techniques, leading to significant reductions in both plausible and correct patch rates when comparing performance on the original and transformed datasets. \thanh{Additionally, we observe notable differences in performance improvements between NPR techniques, suggesting potential biases on NPR evaluation introduced by limited datasets.} Finally, we explore automating the assessment of transformation naturalness by developing a new naturalness metric based on Large Language Models. This metric effectively evaluates naturalness with an AUC of 0.7, offering a promising direction for automating the naturalness assessment of code transformations.

\end{abstract}



\maketitle

\section{Introduction}
~\label{sec:introduction}

Software bug-fixing is a time-consuming, difficult, and tedious task~\cite{winter2022developers}. Hence, Automated Program Repair (APR), which aims to automatically fix bugs in software systems, has emerged as a potential solution to alleviate the onerous burden of this task on software developers. 
Recent years have seen a substantial number of APR techniques~\cite{kim2013automatic, qi2015analysis, le2016history, le2017s3, liu2019tbar, chen2019sequencer, koyuncu2020fixminer, zhu2021syntax, xia2022less, selfAPR2022}, inspiring practical adoption of APR in the industry. 
One notable line of work is known as Neural Program Repair (NPR)~\cite{zhong2022neural, zhong2022standup4npr, zhu2021syntax, ye2022neural}, which leverages advanced Deep Learning (DL) techniques that train on a massive amount of historical data to fix bugs unseen in the past. NPR demonstrates remarkable performance compared to prior approaches, making significant breakthroughs in program repair. 

To evaluate the effectiveness of Neural Program Repair (NPR) systems, prior works~\cite{ye2022neural, xia2022less, zhu2021syntax} have mainly relied on the Defects4J dataset~\cite{just2014defects4j}, which is a widely recognized and high-quality benchmark for Program Repair. Despite its popularity, the Defects4J dataset has two drawbacks. First, it has a limited scale with only 835 bugs, which may not capture the diversity and complexity of real-world bugs in terms of bug types, context, and coding styles. Second, it requires a lot of manual work to add new bugs to the dataset, especially in the stages of bug isolation and verification, resulting in slow updates. For example, Defects4J only increased from 357 to 835 bugs in six years (from 2014 to 2020). These limitations raise concerns about the reliability of Defects4J~\cite{gerobustnpr} for evaluating NPR models as it may struggle to adequately represent real-world bugs.

\thanh{Given these limitations, it is essential to adopt evaluation techniques that extend beyond small-scale, curated benchmarks like Defects4J. In related code domains, common approaches are to conduct either adversarial attacks~\cite{zhou2022natural, zhang2022towards} or metamorphic testing~\cite{rabin2021generalizability, zhang2023challenging} using semantic-preserving transformations. These transformations augment the original benchmark with semantically equivalent but syntactically different code snippets, which are used to evaluate model robustness. However, a critical limitation of these approaches is the lack of assurance that such transformations are naturally induced. For example, Zhou et al.\cite{zhou2022natural} found that adversarial examples generated by MHM\cite{zhang2020generating}, a well-known adversarial attack method, only received an average naturalness rating of only 1.86 out of 5 from human developers. Consequently, these approaches often produce unnatural code transformations that, while effective at exposing model unrobust behaviors, rarely reflect the kinds of transformations encountered in real-world programming as they tend to be awkward or unrealistic~\cite{dyrmishi2023humans, wu2024natural}. As a result, robustness evaluations based on such transformations risk misleading NPR model improvements towards addressing less meaningful unrobust behaviors, which have minimal real-world relevance, thereby hindering long-term progress in enhancing NPR techniques.}

\thanh{In response, we propose shifting the focus towards evaluating NPR techniques' robustness through naturally-occurring data transformations—an approach that better reflects the common, real-world variations developers frequently encounter. This approach, known as \textbf{natural robustness} (or non-adversarial robustness) testing~\cite{wei2023natural, hendrycks2021natural, drenkow2021systematic, gojic2023non}, enables us to prioritize meaningful improvements in NPR’s resilience to natural, real-world data variations before tackling less common adversarial attacks and unnatural transformations~\cite{drenkow2021systematic}. Surprisingly, the current literature has largely overlooked the naturalness aspect when designing and applying such transformations, despite the risk that semantic-preserving transformations may lead to unnatural modifications. We believe this oversight arises from the inherent difficulty of assessing naturalness—even from a human perspective—due to the complexity of defining what is truly 'natural' in the context of code changes.} Different from other domains such as computer vision where the naturalness of data transformation can be quantified by numerical methods such as distance in latent space~\cite{zhao2018generating}, quantifying the naturalness of code transformations is difficult due to the spatial nature of source code~\cite{rabin2021generalizability}. For example, a very small change in Listing~\ref{lst:naming_example} could be unnatural as it contains awkward variable. 

A common approach to assessing the naturalness of code transformations is to rely on manual assessments by human developers~\cite{zhou2022natural, tian2023adversarial}. 
However, these manual assessments are also constrained by the ambiguity surrounding the concept of ``naturalness''. Indeed, our interviews with senior developers, in which we request them to assess the naturalness of example code transformations, reveal that 3 out of 4 participants expressed uncertainty regarding the ``naturalness'' concept. For instance, one developer remarked, ``\textit{the problem could be like I am not sure about what is the naturalness in the natural code transformation. [sic]}''. This ambiguity often leads to confusion among developers when making decisions, limiting the scope of manual naturalness assessments, typically on a small scope, such as assessing the meaning of variable names~\cite{zhou2022natural}.

\begin{lstlisting}[float, language=java, caption=A example of an unnatural bug created by a variable renaming transformation which change variable name val1 to axzt on the bug Time-14 in Defects4J dataset, escapechar=!, label=lst:naming_example]
//Original Code
 public static long safeMultiply(long !\hl{val1}!, int val2) {
    long total = !\hl{val1}! * val2;
    return total;
  }
//Transformed Code
  public static long safeMultiply(long !\sethlcolor{mGreen} \hl{b5saeqw1}!, int val2) {
    long total = !\sethlcolor{mGreen} \hl{b5saeqw1}! * val2;
    return total;
  }
\end{lstlisting}

Motivated by this issue, we are the first to establish concrete criteria for assessing the naturalness of code transformations through interviews with four senior developers.
These assessment criteria enable us to mitigate the ambiguity surrounding the concept of naturalness, facilitating high-quality assessments from human annotators. Next, we aim to assess the naturalness of semantic-preserving transformations created by 18 hand-crafted operators selected from the literature such as method name prediction~\cite{rabin2021generalizability}, code comment generation~\cite{yu2022data}, and code clone detection~\cite{zhang2023challenging}.  
This evaluation enables us to construct a dataset of semantic-preserving transformations annotated with their naturalness, facilitating further analysis of the naturalness of semantic-preserving transformations and revealing insights about their properties and impact. To do this, we survey 10 developers to assess the naturalness of 1098 semantic-preserving transformations, i.e., pairs of original and transformed programs.
These transformations were generated for 220 bugs in the well-known Defects4J~\cite{just2014defects4j} dataset by applying 18 aforementioned hand-crafted operators.

\thanh {We then investigate the impact of unnatural code transformations on the robustness testing of NPR techniques. Specifically, we first assess how the lack of naturalness in the design of semantic-preserving transformations affects their applicability in our benchmark. Next, we evaluate the impact of unnatural code transformations on the outcomes of robustness testing by comparing prediction changes and performance shifts in NPR techniques when exposed to both all semantic-preserving transformations and only natural transformations.} 

Finally, motivated by the high costs of manual naturalness assessments, we conduct an exploratory study to evaluate the feasibility of automating this process. \thanh{We introduce a new naturalness metric, called Relative Naturalness Changes (RNC), which builds on the established Cross-Entropy (CE) metric. Our RNC metric offers two key novelty compared to traditional CE. First, RNC quantifies naturalness by measuring the relative differences in Cross-Entropy values between the original code and the transformed code, thus mitigating the influence of the original code's inherent naturalness on the metric. Second, we leverage large language models (LLMs) to estimate Cross-Entropy, as opposed to traditional n-gram models~\cite{hindle2016naturalness}, enhancing the precision of the CE estimation.}

\begin{longtable}{p{7cm}|p{6cm}}
\caption{Our key findings and implication on Natural Robustness Testing of Neural Program Repair}~\label{tab:key_findings}
\endfirsthead
\endhead
\hline
\rowcolor{lighGray}
\textbf{Findings on Naturalness of semantic-preserving transformation (Section~\ref{sec:rq1})} & \textbf{Implications} \\
\hline
There are only 58.8\% of semantic-preserving transformations that are natural regarding human assessment while the other 19.3\% are deemed unnatural.&  Naturalness assessment should be used to prioritize common, natural code transformations over rare, unnatural ones. \\
\hline
All semantic-preserving transformations include unnatural ones, with the highest unnatural ratio of 62.4\% occurring at the statement level. & Naturalness aspects should be used to guide the design of semantic-preserving transformation, especially at statement-level (1). \\
\hline
\hline
\rowcolor{lighGray}
\textbf{Findings on impact of unnatural semantic-preserving transformation on Robustness Testing of NPR  (Section~\ref{sec:rq2})} & \textbf{Implications} \\
\hline
Unnaturalness in the design of semantic-preserving transformations reduces the applicability of all transformations in natural robustness testing, with the most significant impact occurring at the statement level. & Same as (1) \\
\hline
Only 52.4\% of unrobust inputs are due to natural transformations. Consequently, including all semantic-preserving transformations in robustness testing causes more prediction changes and performance changes than natural transformations alone. Moreover, the relative robustness among NPR techniques varies considerably when evaluated under semantic-preserving transformations with and without naturalness consideration.
& Unnatural code transformations have significant impact on findings of robustness testing, leading to wasted efforts on improving NPR techniques. Thus, natural robustness testing using only natural code transformations should be employed for a more reliable evaluation of NPR techniques.\\
\hline
\hline
\rowcolor{lighGray}
\textbf{Findings on Natural Robustness Testing of NPR (Section~\ref{sec:rq4})} & \textbf{Implications} \\
\hline
NPR techniques exhibit a lack of robustness against natural code transformations, resulting in prediction changes for 8.8\% to 29.4\% of target bugs. 
& Future NPR research should pay more attention on improving robustness of proposed techniques against natural variations of buggy programs, avoiding unrobustness on their predictions.\\
\hline
Prediction changes do not exclusively lead to negative outcomes, they also improve quality of their predictions.
& Future NPR research could explore the characteristics of positive predictions to inform input changes on that improve the quality of generated outputs.\\
\hline
NPR technique performance significantly decline when evaluating on naturally-transformed datasets. While the ranking of NPR techniques mostly remains consistent, significant differences in performance improvements between NPR techniques emerge, highlighting potential biases introduced by evaluation on limited datasets.
& Future NPR research should incorporate natural robustness testing as part of the evaluation process to more reliably assess NPR techniques.\\
\hline
\hline
\rowcolor{lighGray}
\textbf{Finding on Automated Naturalness Assessment of Code Transformations (Section~\ref{sec:rq3})} & \textbf{Implications} \\
\hline
Relying solely on the naturalness of transformed code to assess code transformations can yield inaccurate results due to the strong correlation between the naturalness of the original and transformed code.
& It is essential to develop a new metric that accurately evaluates the naturalness of code transformations by mitigating the influence of the original code's naturalness.\\ \hline
Our proposed metric, Relative Naturalness Changes, calculated by Large Language Models yield promising performance on filtering out unnatural code transformations with AUC of 0.7.
& Automated naturalness assessment of semantic-preserving transformations is worth studying to advance the scalability of natural robustness testing without the high costs associated with manual annotations.\\
\hline
\end{longtable}

\thanh{Table~\ref{tab:key_findings} presents the key findings of our study. Our experimental results revealed significant unnaturalness in the current design of semantic-preserving transformations. This unnaturalness adversely affects robustness testing in two main ways: (1) it limits the applicability of semantic-preserving transformations, particularly at the statement level, and (2) it leads to unreliable evaluations in robustness testing regarding prediction and performance changes in NPR techniques. Moreover, even when using natural robustness testing, we observed that NPR techniques remain unrobust on their predictions, leading to remarkable decline in performance under natural-induced transformations of inputs. However, interestingly, we found that prediction changes do not always result in negative outcomes; they can also enhance the quality of predictions in some cases. Finally, we found that it is feasible to automatically evaluate the naturalness of code transformations using a LLM-based metric, indicating a promising direction for improving the scalability of natural robustness testing.}

\vspace{1mm}

\noindent \textbf{Contributions.} In summary, the main contributions of this work include:

\begin{itemize} 

\item We present a comprehensive set of criteria for evaluating the naturalness of semantic-preserving transformations. To our knowledge, this marks the first attempt to establish such criteria, which provide a foundation for designing and assessing natural semantic-preserving transformations. 

\item We perform a systematic investigation into the unnaturalness of semantic-preserving transformations and their effects on the robustness testing of Neural Program Repair. \thanh{This study covers several aspects, including (1) the applicability of transformations, (2) changes in predictions, and (3) performance variations in NPR techniques.}

\item We are the first to conduct natural robustness testing on Neural Program Repair techniques. This analysis reveals unrobust behaviors and significant performance declines when exposed to naturally-induced transformations of input data. 

\item \thanh{We introduce TransformedDefects4J, a dataset consisting of 1,098 transformations manually labeled for naturalness by human developers. This dataset encompasses a diverse set of 18 transformation operators applied to 220 real-world programs, facilitating future research on (1) assessing the naturalness of semantic-preserving transformations and (2) conducting natural robustness testing of NPR techniques.}

\item We explore automated methods for assessing the naturalness of code transformations using the established naturalness metric, Cross-Entropy (CE), and introduce a novel metric, Relative Naturalness Changes (RNC), based on CE and Large Language Models. This new metric contribute to enhance the scalability of natural robustness testing. 

\end{itemize}

\vspace{2mm}

\noindent \textbf{Data Availability.} To support the open science initiative, we publish our replication package at
\begin{center}
    \url{https://github.com/thanhlecongg/NaturalTransformationForBenchmarkingNPR}
\end{center}

\vspace{2mm}

\noindent \textbf{Research Ethics Statements}. This work involves a two-stage human study with 14 participants. This human study has been approved by the ethics committee of the University of Melbourne under the application reference 2023-26725-41903-5. All participants volunteered for this work and provided their written consent for using their de-identified data in this work. They received a reward of at least \$AUD 23.23 per hour, which follows the Australia National Minimum Wage.

\section{Background and Related Works}
\label{sec:preliminary}

In this section, we briefly introduce Neural Program Repair and related works that empirically assess NPR techniques. Next, we present related works concerning the robustness of AI4Code models. 

\subsection{Program Repair}

\subsubsection{Neural Program Repair} Neural Program Repair (NRP) has achieved impressive performance on program repair~\cite{lutellier2020coconut, xia2022less, zhu2021syntax}. These models, typically based on deep learning models such as transformers, are trained on massive amounts of historical bug-fixes. NPR formulates the bug-fixing process as either (1) a code translation problem, which translates from buggy code into correct code~\cite{chen2019sequencer, lutellier2020coconut, zhu2021syntax, jiang2021cure}, or (2) a cloze, i.e., fill-in-the-blank, task~\cite{xia2022less, xia2023aprllm}, where correct code is synthesized based on the surrounding context. 
Translation-based approaches~\cite{chen2019sequencer, lutellier2020coconut} use deep learning models to learn complex fixing patterns from training data, addressing a broader range of bugs effectively without relying on predefined buggy patterns. 
On the other hand, cloze-based approaches~\cite{xia2022less, xia2023aprllm} leverage Large Language Models (LLMs) to reason about intricate relations between correct fixes and their contexts. 

\subsubsection{Empirical Studies} Neural Program Repair (NPR) still faces its own challenges revealed by prior works~\cite{gerobustnpr, tufano2019empirical, tufano2018empirical, zhong2022neural, zhang2023challenging}, motivating empirical investigations on these systems. Our work closely relates to RobustNPR~\cite{gerobustnpr} which assesses the robustness of NPR techniques against four semantic-preserving transformations. RobustNPR, however, overlooks the naturalness of these transformations, potentially overestimating the lack of robustness in NPR techniques. Additionally, RobustNPR also contains a limited set of four semantic-preserving transformations. Our study extends the efforts of RobustNPR in three dimensions. First, we extend their set of 4 semantic-preserving transformations into an extensive set of 18 transformations. Second, we employ a human study to establish naturalness assessment criteria and assess the naturalness of 1098 transformed bugs generated by these transformations. We also measure the impact of unnatural transformations on the robustness testing of NPR techniques and revisit this evaluation with natural transformations. Finally, we conducted an exploratory study focused on automated naturalness assessment resulting in a new metric for code transformations. These efforts aim to provide reliable and scalable robustness testing of NPR techniques. 

Our work is also related to empirical investigations on NPR techniques and Program Repair in general. Tufano et al.~\cite{tufano2018empirical, tufano2019empirical} empirically explore the ability of Neural Machine Translation (NMT) models in the context of Program Repair. Zhong et al.~\cite{zhong2022standup4npr} found the differences in the setup of NPR techniques and empirically studied these systems in the same setup, namely NPR4J. Less relevant to our work in this paper are empirical and human studies~\cite{smith2015cure, qi2015analysis, liu2020efficiency, le2018overfitting} on Program Repair in general. Smith et al.~\cite{smith2015cure} investigate the overfitting problem with several follow-up works~\cite{qi2015analysis, le2018overfitting, le2019reliability, wang2020automated}. Liu et al.~\cite{liu2020efficiency} evaluate the efficiency of test-based Program Repair. Le-Cong et al. study the memory efficiency of LLM-based Program Repair techniques~\cite{lecong2024flames}. Liu et al.~\cite{liu2019you} explore the impact of fault localization techniques on Program Repair. Kabadi et al.~\cite{kabadifuture} investigate the impact of using future test cases on Program Repair systems. Noller et al.~\cite{noller2022trust} interview software developers to unveil challenges for improving developers' trust in Program Repair systems. Different from these works, our work delves into the problem of the natural robustness of NPR techniques to unveil their true-effectivenss on limited dataset. 

\subsubsection{Program Repair Dataset.} High-quality datasets play a crucial role not only in evaluating the effectiveness of Neural Program Repair systems but also in facilitating fair comparisons between them. Toward these objectives, researchers and practitioners have proposed various benchmarks over time. Just et al.\cite{just2014defects4j} introduced Defects4J\cite{just2014defects4j}, an executable dataset containing 385 real-world bugs designed for benchmarking Program Repair techniques. This dataset has been recently expanded to a second version, including 835 bugs.
Le Goues et al.\cite{le2015manybugs} created two benchmark datasets for C programs, namely ManyBugs and the IntroClass benchmark, which collectively feature 1183 bugs. Following these works, Tan et al.\cite{tan2017codeflaws} introduced CodeFlaws, incorporating 3092 bugs found in student-written programs from Codeforces. Additionally, Madeiral et al.\cite{madeiral2019bears} and Saha et al.\cite{saha2018bugs} introduced Bears and Bugs.jar, containing 251 and 1158 bugs in Java programs.
While the primary focus has been on Java and C programs, researchers have also explored the generalizability of Program Repair techniques by constructing benchmark datasets for other programming languages such as Python~\cite{widyasari2020bugsinpy} and Javascript~\cite{gyimesi2019bugsjs}. Beyond these works, there are other directions aimed at automatically and precisely mining datasets from open-source repositories. These efforts involve detecting bug-fixing commits using related information such as commit messages~\cite{zhou2021spi}, code changes~\cite{nguyen2023multi}, and issue reports~\cite{nguyen2022vulcurator}.

The aforementioned approaches mainly focus on extending Program Repair datasets by incorporating more bugs from diverse resources. This process entails substantial manual effort, including bug isolation and verification, leading to slow progress in dataset updates.
In contrast to these approaches, our study focuses on data augmentation, extending current datasets by introducing new bugs through natural semantic-preserving transformations. This approach complements dataset extension methods, providing an orthogonal approach to increase dataset size. Consequently, it offers a more comprehensive benchmark for evaluating Neural Program Repair systems. 

\subsection{Code Naturalness}
Our work is closely aligned with existing research on code naturalness. The concept was initially proposed by Hindle et al.~\cite{hindle2016naturalness}, who hypothesize that code authored by humans tends to be mostly simple and rather repetitive. This hypothesis was then validated by using n-gram language models to measure the Cross-Entropy of code corpora. Furthermore, this research revealed that software code exhibits even greater levels of repetitiveness compared to natural languages. In subsequent investigations, Campbell~\cite{campbell2014syntax} observed that syntactically incorrect code is less natural than its correct counterparts. Similarly, Ray et al.\cite{ray2016naturalness} found that buggy code has lower naturalness, with its naturalness increasing following the correction of the bug. Building upon this observation, they proposed a naturalness-based static bug finder, which outperforms state-of-the-art techniques such as PMD\cite{copeland2005pmd} or FindBugs~\cite{ayewah2008using}. Khanfir et al.~\cite{khanfir2022codebert} further improve this approach by leveraging the naturalness metric learned from CodeBERT. Additionally, Xia et al.\cite{xia2023aprllm} and Kang et al.\cite{kang2022language} demonstrated that code naturalness can be leveraged to prioritize correct patches over plausible patches in program repair. 

Diverging from these statistical views, our focus shifts towards comprehending naturalness through a human-centric perspective, which introduces unique challenges as discussed in Section~\ref{sec:introduction}. To tackle these challenges, we present the first attempt to establish concrete assessment criteria for the naturalness of code transformations and conduct an extensive human study to evaluate their naturalness. Drawing inspiration from prior research, we also leverage Cross-Entropy acquired from statistical models and Large Language Models to automatically assess code naturalness. However, unlike previous studies, we focus on the naturalness of code transformations that yield unexpected outcomes, as presented in Section~\ref{sec:rq3}, necessitating the introduction of a newly proposed metric, namely Relative Naturalness Changes. In a similar direction to our investigation, ALERT~\cite{zhou2022natural} delves into the naturalness of code transformations. Nevertheless, this approach mainly focuses on variable naming transformations, which only entail the natural substitution of variables. In contrast, our study investigates naturalness generally across program transformations: we consider a broader set of 18 semantic-preserving transformation operators. This expanded scope introduces a unique challenge related to the ambiguity surrounding naturalness, as discussed in Section~\ref{sec:introduction}, mandating the development of a comprehensive methodology to navigate it effectively.

\subsection{Applications of Semantic-preserving Transformation}

\thanh{Our work also aligns with research efforts that apply semantic-preserving transformations to test automated systems in code-related tasks.} The most closely related direction is robustness testing of AI4Code models~\cite{rabin2021generalizability, zhou2022natural, du2023extensive, zhang2023challenging}.  Rabin et al.~\cite{rabin2021generalizability} have proposed a set of 6 operators for generating semantic-preserving transformation. These transformations are then used to evaluate the robustness of AI models for the method name prediction task. Similarly, Zhang et al.~\cite{zhang2023challenging} have designed an extensive set of 15 operators for challenging AI-based clone detection models. Different from these approaches that do not guarantee the naturalness of semantic-preserving transformations, our work investigates the impact of the naturalness of these transformations and proposes concrete criteria for assessing naturalness as well as a new metric for automatically assessing their naturalness. 

\thanh{Less relevant to our works are recent efforts on adversarial attacks~\cite{zhou2022natural, zhang2022towards, tian2023adversarial} and program analysis testing~\cite{zhang2023statfier, wu2024natural}. These approaches, however, focus on transformations that \textit{intentionally} exploit edge cases that cannot be effectively handled by automated techniques such as AI-driven vulnerability detection~\cite{zhou2022natural} or static analyzers~\cite{zhang2023statfier}. Different from these approaches, our approach focuses on transformations that \textit{naturally} occur to unveil the true effectiveness of NPR techniques as discussed in Section~\ref{sec:introduction}}. 
\begin{table*}[t]
\caption{Basic information of interviewees. "ID," "Exp," and "Roles" present a participants identification within our study, years of experience till the time of the interview, and current and past roles.}
\label{tab:inverview_participant_info}
\begin{tabular}{l|l|l|l}
\toprule
\textbf{Phase}                      & \textbf{ID} & \textbf{YoE} & \textbf{Roles}                                                                                                                                    \\ \hline
\multirow{4}{*}{Interview} 
                           & I1                & 6                   & Software Developer at Reputed Technology Company A                                                                                        \\ \cline{2-4} 
                           & I2                & 8                   & \begin{tabular}[c]{@{}l@{}}Researcher at Reputed University B\\ Former Project Manager and Software Developer at Reputed Bank C\end{tabular} \\ \cline{2-4} 
                           & I3                & 12                  & \begin{tabular}[c]{@{}l@{}}Researcher at Reputed University B\\ Former Software Developer at Reputed IT Company D\end{tabular}       \\ \cline{2-4} 
                           & I4                & 22                  & Senior Engineer at Reputed Bank E
                           \\
                           \bottomrule
\end{tabular}
\end{table*}

\section{Assessment Criteria for Naturalness of Code Transformations}~\label{sec:criteria}
As indicated in Section~\ref{sec:introduction}, the ambiguity surrounding the concept of 'naturalness' in code transformations presents a significant challenge in evaluating their naturalness. This motivates the following research question:
\begin{center}
\textbf{RQ1: What are criteria for assessing naturalness of code transformations?}
\end{center}
To address this question, we conduct interviews with senior developers to gather their insights. In this section, we outline our interview design followed by our data analysis approach. Finally, we present our findings and establish concrete assessment criteria.

\subsection{Interview Design}

\subsubsection{Participant Recruitment.} We initially contacted potential participants through emails and subsequently broadened our participant pool using a snowball process~\cite{goodman1961snowball}. In this process, existing participants suggested additional individuals who might be interested in our research. Participants were required to have a minimum of 5 years of experience for the interview phase. \thanh{Note that, to avoid potential bias, we invited independent participants which did not involve as authors of our papers.}
Overall, our study included 4 senior developers and researchers in the interview Table~\ref{tab:inverview_participant_info}. 

\subsubsection{Interview.}\label{sec:interview} All interviews were conducted via Zoom and recorded with the permission of the participants. These records are de-identified and transcribed into text for further analysis. The average and standard deviation of the interview were 25.3 and 5.4 minutes, respectively. The interviews were semi-structured and divided into three parts.

\vspace{2mm}

\textbf{Part 1:} We provided a brief introduction to the interviewees regarding our research project and the objectives of our interview. Moreover, we engaged them with a set of questions regarding robustness and naturalness. These questions aimed to enhance their comprehension, while also gathering insights from their perspective.

\vspace{2mm}

\textbf{Part 2:} We posed open-ended questions to gain insight into interviewees' perspectives regarding the naturalness of code transformations. In particular, we asked about the properties they consider related to the naturalness of these transformations. For each identified property, we encouraged them to provide illustrative examples and brief rationales for their selections. Additionally, we also validate these properties by seeking affirmation from interviewees regarding the properties mentioned by their peers in earlier interviews. Finally, we asked interviewees about which context and inputs are necessary for them to assess the naturalness of code transformations.

\vspace{2mm}

\textbf{Part 3:} We asked about the demographic aspects of the interviewees. Specifically, we collected information about their past experience, including their employment roles, their organizations, and years of experience.

\subsubsection{Data Analysis}
Next, for establishing assessment criteria, we qualitatively analyze the aforementioned interviews to derive a set of criteria for the naturalness assessment of semantic-preserving transformations as follows:
\begin{itemize}
    \item \textbf{Thematic Analysis.} First, we manually analyze and conduct Thematic Analysis~\cite{braun2006using} to obtain themes in interviewees' response. We consider these themes as properties, which are considered to be related to the naturalness of code transformations. \thanh{Two authors of the paper were involved in the Themantic Analysis. The disagreements were resolved through a discussion between two authors in which one of the authors serve as moderator.}

    \item \textbf{Open Card Sorting.} Then, we employed an Open Card Sorting~\cite{spencer2009card} discussion between two authors to categorize themes obtained from Thematic Analysis into main themes. 
    \item \textbf{Criteria Analysis.} Finally, we manually analyze interviewees' responses regarding these main themes to derive naturalness assessment criteria. 
\end{itemize}

\begin{table}[]
\caption{Themes, i.e., Properties, obtained from interviewees' responsens}~\label{tab:theme}
\begin{tabular}{l|l}
\toprule
\textbf{Themes}                                                               & \textbf{Main Themes}   \\
\hline
Context-appropriation of   variable names                                     & Code Convention        \\
Code convention                                                               & Code Convention        \\
Code modularity                                                               & Code Readability       \\
Code readability                                                              & Code Readability       \\
Code complexity                                                               & Code Readability       \\
Code style                                                                    & Code Convention        \\
Appropriation of identifiers  toward the problem/domain & Code Convention        \\
Consistency between structure name and its content.                         & Code Convention        \\
Complexity of structures                                                      & Code Readability       \\
Co-occurrences of code elements.                                             & Code Convention        \\
Naming convention                                                             & Code Convention        \\
Meaning of variable names                                                     & Code Readability       \\
Control-flow complexity of code                                                & Code Readability       \\
Consistency in terms of dependency.                                          & Appropriateness   \\
Runtime resource usage                                                        & Runtime resource usage \\
Project-specific conventions                                                  & Code Convention        \\
Comment Style                                                                   & Code Convention        \\
Domain/paradigm-appropriation of naming and code structure.                 & Code Convention        \\
Method signature and docstring   convention                                   & Code Convention        \\
Complexity of variable name                                                   & Code Readability    \\
\bottomrule
\end{tabular}
\end{table}

\subsection{Findings}
Through the Thematic Analysis, we obtain 20 themes which are presented in Table~\ref{tab:theme}. Then, we categorize these themes into 4 main themes including:
\begin{itemize}
    \item Coding Convention
    \item Code Readability
    \item Appropriateness
    \item Runtime Resource Usage
    
\end{itemize}
To ensure the reliability of this discussion, we employed Cohen's Kappa~\cite{cohen1960coefficient}, a well-known inter-rater agreement measurement. The overall Kappa value is 0.75, which indicates substantial agreement between the two authors. From these themes, we decided to select the first two, Coding Convention and Code Readability as they reached a consensus among all interview participants. The latter two themes were excluded as they were mentioned by only one interviewee. 

Then, by further analyzing interviews, we found that the developer will consider a code transformation as unnatural if it reduces the code readability of the original code. For example, an interviewee mentioned that ``\textit{I guess the essential point is still human readability. Right? If it is very easy to be understood by human, then that's natural. And it's also good for code interpretation. If it is very complex and human. Not normally. It's very difficult for humans to understand is intention. That's bad, and it's unnatural.[sic]}''. We also found that a code transformation is also deemed unnatural by developers if it breaks the coding convention of the original code. For example, an interviewee mentioned that ``\textit{the most important part here is to make sure the code is consistent with the rest of the code. }''
Given these insights, we define two criteria for assessing the naturalness of semantic-preserving transformation as follows:
\begin{tcolorbox}
\begin{definition}\label{def:criteria}
    (Naturalness Assessment Criteria) A semantic-preserving transformation is deemed unnatural if at least one of the following conditions is met:
    \begin{enumerate}
        \item It reduces the code readability of the original code;
        \item It breaks the coding convention of the original code.
    \end{enumerate}
\end{definition}
\end{tcolorbox}
\section{Manual Assessment and Empirical Study Design}
\label{sec:design}

In subsequent investigations, to understand the impact of naturalness on the evaluation of NPR techniques and the actual robustness of these systems, we conduct manual assessment and empirical studies to answer the following three research questions:

\vspace{2mm}

\noindent\textbf{RQ2: How natural are semantic-preserving code transformations?} Semantic-preserving transformations are frequently employed to augment coding benchmarks and assess the robustness of AI models for code~\cite{zhuo2023data}. However, the extent to which these transformations preserve the naturalness of the code remains largely unexplored. Notably, developers may perceive semantic-preserving transformations as unnatural. Consequently, these transformations can generate unnatural programs, leading to false alarms regarding the robustness of Neural Program Repair systems. To understand this issue, this research question investigates the naturalness of 1098 semantic-preserving transformations within the context of software bug-fixing. 

\vspace{2mm}

\noindent\textbf{RQ3: What is the impact of unnatural semantic-preserving transformation on the evaluation of Neural Program Repair systems?}As discussed in the previous research question, semantic-preserving transformations can introduce unnatural elements, triggering rare and less meaningful unrobust behaviors in NPR techniques. These behaviors often require unnecessary effort to address. This research question aims to assess the impact of such unnatural semantic-preserving transformations on the evaluation of Neural Program Repair systems, focusing on three key aspects: (1) the applicability of transformations, (2) changes in prediction, and (3) performance variations. 

\vspace{2mm}

\noindent\textbf{RQ4: How robust are Neural Program Repair systems against natural code transformations?} This research question revisit robustness of 7 well-known state-of-the-art Neural Program Repair systems including SequenceR~\cite{chen2019sequencer}, Recoder~\cite{zhu2021syntax}, RewardRepair~\cite{ye2022neural}, SelfAPR~\cite{selfAPR2022}, AlphaRepair~\cite{xia2022less}, Incoder~\cite{jiang2023impact} and RepairLLama~\cite{silva2023repairllama}. This investigation seeks to provide insights about the true effectiveness of NPR techniques againts natural variations of input data.  

\vspace{2mm}

\noindent\textbf{RQ5: Is it feasible to automate naturalness assessment of semantic-preserving transformations using existing naturalness metrics?} While manual assessment of the naturalness of code transformations is reliable, as we will show in Section~\ref{sec:rq1}, the process is time-consuming and costly. In our study, for instance, an assessment of 1098 transformations costs approximately 60 working hours and around 800 US dollars. The costs pose challenges in ensuring the scalability of robustness evaluation. To shed light on this challenge, this research aims to explore the feasibility of automatically assessing naturalness of code transformations using existing naturalness metric, i.e., Cross-Entropy~\cite{hindle2016naturalness} calculated with various statistical language models including n-gram~\cite{hindle2016naturalness}, GPT-Neo~\cite{gptneo}, BLOOM~\cite{scao2022bloom} and Code Llama~\cite{roziere2023code}.

\begin{figure}
    \centering
    \includegraphics[width=\textwidth]{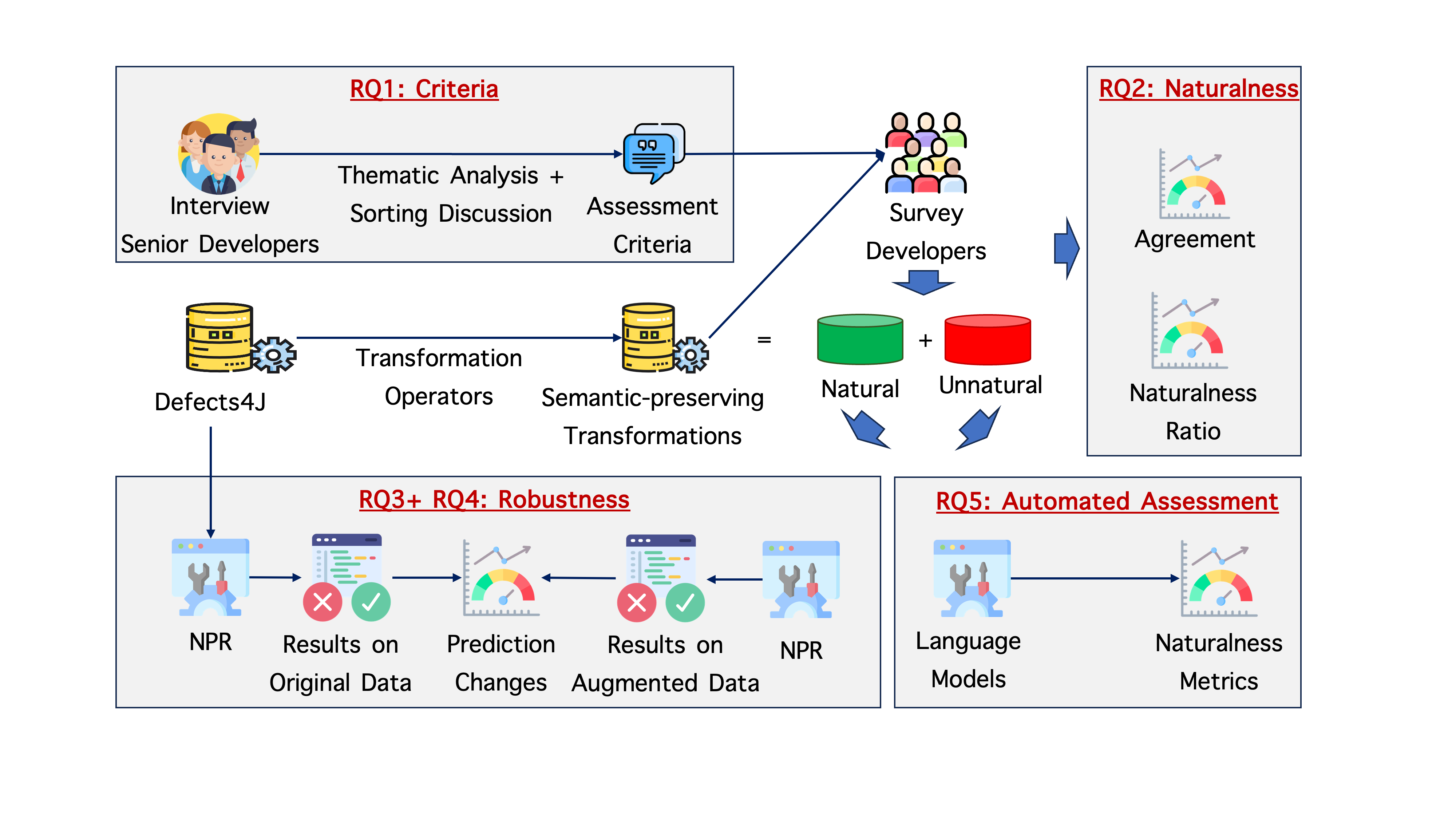}
    \caption{Overview of our study design}
    \label{fig:overview}
\end{figure}

\vspace{2mm}

To address these research questions, our research methodology follows a mixed qualitative and quantitative approach~\cite{easterbrook2008selecting} as illustrated in Figure ~\ref{fig:overview}. Particularly, we begin with an interview of senior developers, presented in Section~\ref{sec:criteria} for establishing a set of specific criteria that can be used to assess the naturalness assessment of code transformations (RQ1). It is followed by a user study stage designed to identify natural code transformations as assessed by human annotators. Annotations collected during the user study will be analyzed quantitatively to measure the ratio of unnatural semantic-preserving transformations (RQ2) and their impact on the evaluation of Neural Program Repair (RQ3). Subsequently, we revisit the robustness of Neural Program Repair against natural semantic-preserving transformations (RQ4). Finally, we explored the feasibility of automatically identifying natural code transformations using the existing naturalness metric, i.e., Cross-Entropy and statistical language models (RQ5). 

\subsection{Manual Assessment}
In this section, we provide an overview of our manual assessment process for evaluating the naturalness of semantic-preserving transformations, carried out through an online survey.

\subsubsection{Participant Recruitment.} Similar to the interview, we also use snowball process~\cite{goodman1961snowball} to recruit participants for survey.  Participants were required to have a minimum of 1 years of experience with Java which is the targeted programming language in this study. \thanh{These participants were independent of the authors and were not involved in other parts of this study.} Overall, our study involve 10 developers in the survey stage. Demographic information about these participants can be found in Table~\ref{tab:participant_info}. 

\begin{table*}[t]
\caption{Basic information of survey participants. "ID," "Exp," and "Roles" present a participants identification within our study, years of experience till the time of the interview, and current and past roles.}
\label{tab:participant_info}
\begin{tabular}{l|l|l|l}
\toprule
\textbf{Phase}                      & \textbf{ID} & \textbf{YoE} & \textbf{Roles}         \\                                                             \hline
\multirow{5}{*}{Survey}    
& S1                &  1                   & Research Engineer at Reputed University H                     \\ \cline{2-4} 
                           & S2                &  1                  &  Software Developer at Software Development Company F                      \\\cline{2-4}  
                           & S3                &  1                   & Software Developer at Software Development Company G                       \\\cline{2-4} 
                            & S4                &  1                   & Research Engineer at Reputed University H                       \\ \cline{2-4}
                            & S5                &  2                   & Software Developer at Software Development Company I                                                                                                                                   \\ \cline{2-4} 
                           & S6                &  3                   & Software Developer at Bank K                              \\ \cline{2-4} 
                           & S7                &  3                  &  Software Developer at Software Development Company L      \\ \cline{2-4} 
                           & S8                &  4                   &  Software Developer at Search Engine Provider M                      \\ \cline{2-4}  
                           & S9                &  6                   & Software Developer at Bank N                       \\\cline{2-4}  
                           & S10                &  7                   & Software Engineer at Social Network Company P                      \\  \bottomrule
                           
\end{tabular}
\end{table*}

\subsubsection{User Study.}~\label{sec:user_study} Next, we conducted a user study involving 10 professional developers on assessing of the naturalness of semantic-preserving transformations. Each code transformation is assessed by 5 out of 10 developers following prior work in Program Repair~\cite{le2019reliability}. In the subsequent paragraphs, we provide comprehensive details about the user study, including task design, context information and data collection and processing.

\vspace{2mm}

\textbf{Task Design.} To assessment of semantic-preserving transformations, we follow assessment criteria established in Definition~\ref{def:criteria}. Particularly, to validate the criteria, we ask participants the following questions:
\begin{itemize}
    \item "Does the code transformation reduce code readability?", validating the first condition;
    \item "Does the code transformation break the coding conventions?", validating the second condition.
\end{itemize}
Participants were asked to provide scores using a 4-point Likert scale~\cite{joshi2015likert}, where 1, 2, 3, and 4 indicate Disagree, Weakly Disagree, Weakly Agree, and Agree, respectively.

\begin{figure}[t]
    \centering
    \fbox{\includegraphics[width=\textwidth]{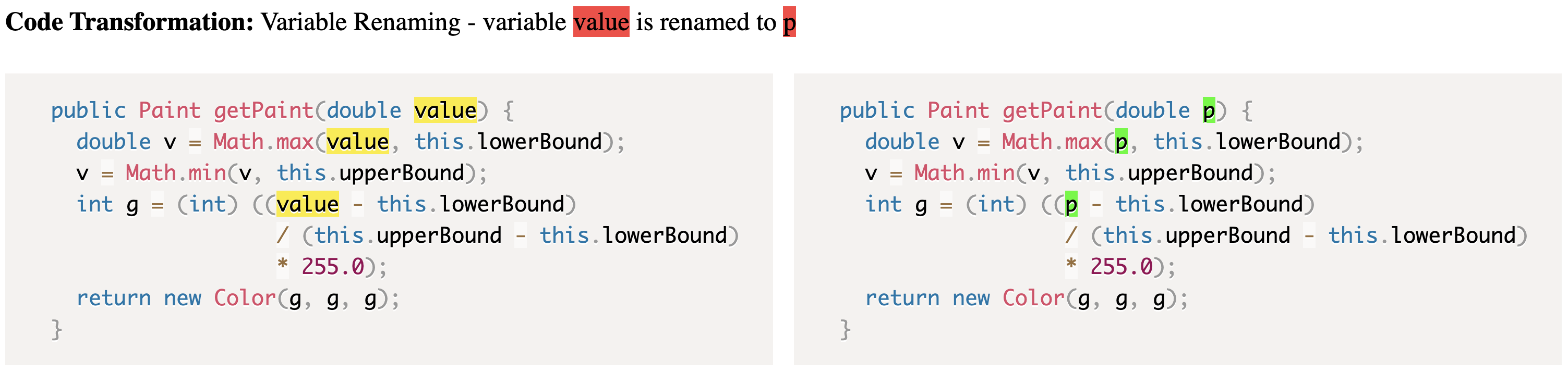}}
    \caption{An example of survey for user study}
    ~\label{fig:survey_eg}
    \label{fig:enter-label}
\end{figure}

\vspace{2mm}

\textbf{Context.} To ensure participants had sufficient information for their assessments, we first provided them with an assessment guideline~\footnote{Please see our replication package~\cite{package} for a sample of this guideline}. This guideline briefly explains our research topic including a summarization of semantic-preserving transformation operators with examples and what participants need to do to complete the tasks at the start of experiments. Afterward, participants can complete their assessment one by one through surveys hosted by Qualtrics~\footnote{https://www.qualtrics.com/}. Figure~\ref{fig:survey_eg} shows an example of information we provide to participants for their assessment. For each assessment of a code transformation, we provide a concise summary of this transformation and a visual representation of the difference between the original code and the transformed code method. As suggested by interviewees (Section~\ref{sec:interview}), we also provide the surrounding context, including the associated method, file, and project. 

\vspace{2mm}

\textbf{Data Collection and Processing.} We begin with de-identifing Qualtrics' response of participants to align with research ethics. Next, to provide more precise assessment from these surveys, we processed data collected from participants' responses in our human study. Regarding the naturalness assessment, we process the responses from participants to derive the final assessment using our established criteria for naturalness evaluation (see details in Definition~\ref{def:criteria}). Specifically, we determine that a code transformation is considered unnatural if a participant provides a positive response, i.e., weakly agree or agree, for at least one out of two conditions, i.e., code readability and coding convention. Regarding completion time, we observed that there are outlier completion times in our records. Hence, we have randomly checked with a few participants and found that these outliers happen due to network connection issues as our surveys are distributed online. To address this, we have mitigated these outliers by implementing a cutoff at the 99th percentile of completion times.
Particularly, we exclude all completion times which were in top 1\%.
\vspace{2mm}

\subsection{Dataset Constructions}
\subsubsection{Bug Dataset.} Since our primary focus is on Automated Program Repair (APR), we employed the Defects4J dataset version 2.0.0, which is a widely used benchmark by the APR community. This dataset contains 835 bugs collected from 17 open-source projects. In alignment with the state-of-the-art Neural Program Repair techniques~\cite{chen2019sequencer, ye2022neural, selfAPR2022}, which mainly focus on single-hunk bugs, we also restricted our dataset to this specific category of bugs within Defects4J. Consequently, we identified and retained 334 out of the 835 bugs. After that, we filtered out bugs that are not applicable to apply semantic-preserving code transformations (see details of these code transformations in the subsequent paragraphs). For instance, we excluded bugs in which the buggy lines do not belong to any methods, as our scope on code transformations requires the presence of the corresponding buggy method. Finally, we obtained a final dataset of 225 bugs. A full list of the bugs used in our study can be found in our replication package~\cite{package}.

\vspace{2mm}

\subsubsection{Scope of Code Transformations.} The scope refers to \textit{the specific regions within the original buggy code where code transformations are applied}. Given the diverse types of input scope accepted by Neural Program Repair techniques, such as the entire buggy method~\cite{zhu2021syntax} or the buggy line along with its surrounding context~\cite{ye2022neural}, it is essential to define a common scope that is applicable to the majority of these techniques. After careful consideration, we have decided to limit our scope of code transformations to buggy methods. This choice aligns with the input scope commonly accepted by most Neural Program Repair techniques~\cite{chen2019sequencer, zhu2021syntax, selfAPR2022, ye2022neural, xia2022less}. Additionally, we also focus on semantic-preserving transformations that affect  buggy locations as these transformations have been shown to have a more significant impact on Neural Program Repair techniques, as demonstrated by Ge et al.~\cite{gerobustnpr}.

\vspace{2mm}

\begin{table}[t]
    \centering
    \caption{The descriptions and examples of variable renaming (naming-level) transformation}
    \label{tab:naming}
    \resizebox{0.75\textwidth}{!}{
    \begin{tabular}{l l l l}
    \hline
        \textbf{Operator} & \textbf{Description} & \textbf{Original} & \textbf{Transformed} \\ \hline
        \textit{RenameVariable-1~\cite{gerobustnpr}} & Replace a variable name  & int data; & int d; \\ 
        ~ & by its first character & data = 1; & d = 1; \\ \hline
        \textit{RenameVariable-2~\cite{zhou2022natural}} & Replace a variable name by & int list; & int arr; \\ 
        ~ & substitutions from CodeBERT~\cite{zhangyin2020codebert} & list = [1, 2]; & arr = [1, 2]; \\ \hline
    \end{tabular}
    }
\end{table}

\begin{table}[t]
    \centering
    \caption{The descriptions and examples of expression-level code transformations}~\label{tab:expression}
    \resizebox{0.85\textwidth}{!}{
    \begin{tabular}{ l l l l }
    \hline
        \textbf{Operator} & \textbf{Description} & \textbf{Original} & \textbf{Transformed} \\ \hline
        \textit{SwitchRelation} & Transform Relational & a < b & b > a \\ 
        ~ & Expressions & ~ & ~ \\ \hline
        \textit{Unary2Add} & Modifications to unary  & i++; & i = i + 1; \\ 
        ~ & operations/Increments & ~ & i += 1; \\ \hline
        \textit{Add2Equal} & Convert add/subtract & a += 9; & a = a + 9; \\ 
        ~ & assignments to equal & b -= 10; & b = b - 10; \\ 
        ~ & assignments & Merge variable declarations & ~ \\ \hline
        \textit{MergeVarDecl} & ~ & int a; & int a, b; \\
        ~ & ~ & int b; & ~ \\ \hline
        \textit{InfixDividing} & Divide an in/pre/post-fix  & x = a + b * c & temp = b * c; \\
        ~ & expression into two expressions  & ~ & x = a + temp \\ 
        ~ & whose values are stored & ~ & ~ \\ 
        ~ & in temporary variable & ~ & ~ \\ \hline
        \textit{SwitchEqualExp} & Switch the two expressions on & a == b & b == a \\ 
        ~ & both sides of the infix expression & ~ & ~ \\ 
        ~ &  whose operator is = & ~ & ~ \\ 
        ~ &  & ~ & ~ \\ \hline
        \textit{SwitchStringEqual} & Switch the string equals. & a.equals(b) & b.equals(b) \\ \hline
    \end{tabular}
    }
\end{table}

\begin{table}[t]
    \centering
    \caption{The descriptions and examples of statement-level code transformations}
    \label{tab:statement}
    \resizebox{0.85\textwidth}{!}{
        \begin{tabular}{ l l l l }
    \hline
        \textbf{Operator} & \textbf{Description} & \textbf{Original} & \textbf{Transformed} \\ \hline
        \textit{For2While} & Transform for loop-to & for(i=1; i<10; i++)\{ & i = 0; \\ 
        ~ & while-loop & //Body & While (i<10)\{ \\ 
        ~ & ~ & \} & //Body \\ 
        ~ & ~ & ~ & i ++; \} \\ \hline
        \textit{While2For} & Transform while-loop & i = 0; & for(i=1; i<10; i++)\{ \\ 
        ~ & to for-loop & While (i<10)\{ & //Body \\ 
        ~ & ~ & //Body & \} \\ 
        ~ & ~ & i ++;\} & ~ \\ \hline
        \textit{ElseIf2If} & Transform If…Else & if (a < 80)\{ & if (a < 80)\{ \\ 
        ~ & if … to If…Else… & //BodyA & //BodyA \\ 
        ~ & ~ & \} else if (a < 100)\{ & \} else \{ \\ 
        ~ & ~ & //BodyB & if (a < 100)\{ \\ 
        ~ & ~ & \} else \{ & //BodyB \\ 
        ~ & ~ & //BodyC & \} else \{ \\
        ~ & ~ & \} & //BodyC \}\} \\\hline
        \textit{Switch2If} & Transform Switch-Case & switch (a)\{ & if (a == 60)\{ \\ 
        ~ & to If-Else  & case 60: //Body A & //Body A \\ 
        ~ & ~ &  default: //Body B & \} else \{ \\ 
        ~ & ~ & \} & //Body B\} \\ \hline
        \textit{SwapStatement} & Swap two statement & a = b + 10; & c = d + 1; \\ 
        ~ & without control and & c = d + 1; & a = b + 10; \\
        ~ & data dependency & ~ & ~ \\ \hline
        \textit{ReverseIf} & Switch two code blocks  & if (condition)\{ & if (!condition)\{ \\ 
        ~ & in the if statement  & //BodyA & //BodyB \\ 
        ~ & and the corresponding & \} else \{ & \} else \{ \\ 
        ~ & else statement & //BodyB\} & //BodyA\} \\\hline
        \textit{If2CondExp} & Change a single & if (condition)\{ & condition ? \\ 
        ~ & if statement into  & //StatementA & StatementA \\ 
        ~ & a conditional expression & \} else \{ & : StatementB \\ 
        ~ & statement & //StatementB \} & ~ \\ \hline
        \textit{ConfExp2If} & Change a conditional & condition ?  & if (condition)\{ \\ 
        ~ & expression statement into & StatementA & //StatementA \\ 
        ~ & a single if statement & : StatementB & \} else \{ \\ 
        ~ & ~ & ~ & //StatementB\} \\\hline
        \textit{DividingComposedIf} & Divide a if statement  & if (condition1 \&\& & if (condition1)\{ \\
        ~ & with a compound condition  & condition2)\{ &if (condition 2)\{ \\
        ~ & ($\wedge, \vee, \neg$) into two & //Body \} & //Body \\
        ~ & nested if-statements & ~ & \} \} \\ \hline
    \end{tabular}
    }
\end{table}

\subsubsection{Code Transformation Operators.} To generate semantic-preserving transformations, a common approach is to utilize hand-crafted operators~\cite{zhang2023challenging, rabin2021generalizability}. In this study, we curated these semantic-preserving transformation operators from the literature~\cite{rabin2021generalizability, zhang2023challenging, yu2022data}. This process yielded a total of 6, 15, and 17 operators from Rabin et al.~\cite{rabin2021generalizability} on method name prediction, Zhang et al.~\cite{zhang2023challenging} on code clone detection, and Yu et al.~\cite{yu2022data} on code comment generation. After eliminating any duplicated operators, we obtained a total of 27 unique operators. We subsequently exclude 4 operators that are deemed unsuitable or challenging to ensure semantic equivalence within our scope. For example, we have excluded the "Unused Statement" transformation operator proposed by Rabin et al.~\cite{rabin2021generalizability}, which involves the insertion of an unused string declaration into a randomly selected basic block within a method. We made this exclusion because we only focus on code transformations that affect buggy lines as mentioned in our previous discussion regarding scope of code Transformations. Additionally, we filtered out 6 operators that were not applicable to any of the bugs within our bug dataset. \thanh{These transformations include:
\begin{itemize}
\item  Do2While: Converts `do-while' loops into `while' loops.
\item  IfElse2IfElseIf: Converts `if-else' statements with an `if' inside the `else' block into `if-elseif' statements.
\item  ModifyConstant: Modifies constant declarations.
\item  DivideVarDecl: Splits compound variable declarations into separate statements.
\item  DividePrePostFix: Separates prefix and postfix expressions into distinct sub-statements.
\item LoopIfContinue2Else: Converts `if' statements with `continue' inside loops into if-else statements
\end{itemize}}

As a result, we employed a final set of 18 semantic-preserving transformation operators for the next steps in our study. These operators were further categorized into three distinct levels: (1) naming-level transformations, (2) expression-level transformations, and (3) statement-level transformations, each of which denotes the specific level where these transformations occur. A full list of these operators are presented in Tables~\ref{tab:naming}, ~\ref{tab:expression}, ~\ref{tab:statement}.

\begin{table}[]
    \centering
    \caption{Statistics of semantic-preserving transformations per category.}
    \label{tab:stats_transform}
    \begin{tabular}{c|ccc|c}
        \hline
        \textbf{Categories} & \textbf{Naming} & \textbf{Expression} & \textbf{Statement} & \textbf{Total} \\
        \hline

        \textbf{Number of Transformations}  & 906 & 155 & 117 & 1098 \\
        \hline

    \end{tabular}
\end{table}

\begin{table}[]
    \centering
    \caption{Statistics of semantic-preserving transformations per bug.}
    \label{tab:stats_transform_per_bug}
    \begin{tabular}{c|ccc}
        \hline
        \textbf{} & \textbf{Minimum} & \textbf{Maximum} & \textbf{Average} \\
        \hline

        \textbf{Number of Transformations}  & 1 & 12 & 5.23 \\
        \hline

    \end{tabular}
\end{table}

\thanh{\textbf{Correctness Validation.} To validate the correctness of programs transformed by code transformation operators, we employed a three-step validation process. First, we selected transformation operators from prior work that were explicitly designed to preserve semantics. Second, we enforced additional constraints on the transformations to prevent naming conflicts, discarding any transformations that resulted in invalid identifiers. Finally, we executed the Defects4J test suite to verify that the transformed programs were not only compilable and executable but also retained the original program behavior, as reflected by the same test case outcomes (i.e., passing or failing).}

\subsubsection{Data Statistics.}
Finally, we obtain 1098 code transformations by applying these 18 transformation operators on 220 bugs in Defects4J, resulting in a new dataset of 1318 bugs. Table~\ref{tab:stats_transform} and ~\ref{tab:stats_transform_per_bug} shows statistics of semantic-preserving transformations per category and per bug, respectively. Among the three categories, naming-level operators exhibit the highest coverage with 854 transformations, followed by expression-level and statement-level operators with 104 and 140 transformations, respectively. These statistics are understandable, as most bugs involve variable names, making them suitable candidates for naming-level operators. Conversely, expression-level and statement-level operators necessitate stricter conditions for application. Regarding the number of transformations per bug, each bug contains a maximum of 12 transformations, with the majority experiencing 3-6 transformations and an average of 5.23. This distribution ensures that our experimental results will not be biased by a subset of bugs in Defects4J.


\subsection{Experimental Settings for Neural Program Repair}
\subsubsection{NPR System Selection} Ideally, we want to consider all state-of-the-art Neural Program Repair (NPR) techniques in our study. Unfortunately, we are limited by the fact that many NPR techniques either provide incomplete replication packages or are too hard to execute, replicate, or extend. Therefore, inspired by Liu el al.~\cite{liu2020efficiency}, we select NPR techniques based on 3 following criteria:
\begin{itemize}
    \item \textbf{Availability}: This criterion concerns the availability of a replication package for NPR techniques. Note that, as NPR techniques require an expensive training process, this criterion also involves the availability of trained models;
    \item \textbf{Configurability}: The NPR techniques require extensive configurations of environments including dependency versions. We exclude NPR techniques with insufficient configuration information;
    \item \textbf{Executability}: Finally, our selection ensures that we can successfully execute NPR techniques and obtain similar results on 20 random bugs with their original experimental settings.
\end{itemize}

\begin{table}[t]
\caption{Included and excluded NPR tools for our study.}~\label{tab:npr}
\resizebox{0.85\textwidth}{!}{
\begin{tabular}{l|l|l}
\toprule
\textbf{Selected}? & \textbf{Reason}          & \textbf{NPR techniques} \\ \hline
No       & Code Availability    & ~\cite{tang2021grammar}, Fix-Filter-Fix~\cite{hong2021fix},~\cite{connor2022can} \\ \hline
No       & Model Availability    & Edits~\cite{ding2020patching}, CoCoNUT~\cite{lutellier2020coconut}, CODIT~\cite{chakraborty2020codit}         \\
~ & ~ &DEAR~\cite{li2022dear}, DLFix~\cite{li2020dlfix}, KNOD~\cite{jiang2023knod}, Codex~\cite{xia2023practical}~\tablefootnote{For this work, we consider CodeX as a representative LLM due to its remarkable performance surpassing NPR techniques, while other LLMs under-performed existing techniques. Unfortunately, CodeX was shut down by OpenAI in March 2023, making it unavailable for our study.}          \\ \hline

No       & Configurability & GrasP~\cite{tang2021grasp}            \\ \hline
No       & Executability   & CURE~\cite{jiang2021cure}~\tablefootnote{We were able to successfully execute and replicate their results on the QuixBugs dataset, following the instructions provided by the authors. Unfortunately, the lack of instructions regarding the Defects4J dataset, which is our focus, leads to our failure to replicate their results. While we have attempted to reach out to the authors through GitHub issues, we have not yet received a response. Consequently, due to time constraints, we have decided to exclude this tool from our study and consider it as having executability issues.}            \\ \hline
Yes      &                 & SequenceR~\cite{chen2019sequencer}, RewardRepair~\cite{ye2022neural},SelfAPR~\cite{selfAPR2022}         \\ 
~ & ~ & AlphaRepair~\cite{xia2022less}, Recoder~\cite{zhu2021syntax}, Incoder~\cite{jiang2023impact}, RepairLlama~\cite{silva2023repairllama}          \\ \bottomrule
\end{tabular}
}
\end{table}

We conducted a comprehensive literature review to collect recent state-of-the-art NPR techniques. As of February 2023, we identified 19 NPR techniques. We then rigorously assessed these tools against the aforementioned criteria and found that 7 out of 19 NPR techniques met our requirements including SequenceR~\cite{chen2019sequencer}, Recoder~\cite{zhu2021syntax}, RewardRepair~\cite{ye2022neural}, SelfAPR~\cite{selfAPR2022}, and AlphaRepair~\cite{xia2022less}. A full list of identified tools and our selection process is presented in Table~\ref{tab:npr}.

\subsubsection{Experimental Settings} In this section, we present an overview of experimental settings for Neural Program Repair techniques in our study.

\vspace{2mm}

\textbf{Implementation Details.} \thanh{To evaluate selected NPR techniques on bugs from code transformations, we collect the relevant replication packages, including inference and validation scripts. We then integrate deep learning and LLM-based approaches into the Cerberus and FLAMES frameworks developed by Shariffdeen et al.\cite{cerberus} and Le-Cong et al.\cite{lecong2024flames}, respectively. DL model experiments run on an Nvidia RTX A5000 GPU with 16 GB VRAM and an Intel i7-10700K CPU with 64 GB RAM, while LLM experiments use an NVIDIA A100 GPU with 80 GB VRAM, 250 GB RAM, and a 32-core Intel Xeon CPU.}
Regarding the configuration of repair tools, following prior works~\cite{chen2019sequencer, selfAPR2022}, we set a beam width of 50 for NPR techniques except SelfAPR and AlphaRepair. For SelfAPR,  10 distinct models, we collected the top 5 patches from each model to obtain the top 50 patches. As for AlphaRepair, due to its extensive use of templates, a beam width of 50 resulted in a substantial number of generated patches (in the thousands). Consequently, we decided to set its beam width to 5 to ensure a reasonable number of patches (around 50 to more than 200) for a manageable generation time. \thanh{For LLM-based Program Repair approaches, we set the beam width to 10, following the default parameters, due to resource constraints and the substantial VRAM consumption of these models~\cite{lecong2024flames}.}

\vspace{2mm}

\textbf{Fault Localization.} As the performance of Automated Program Repair techniques may be affected by fault localization information, we have standardized the configuration of all selected NPR techniques to use the same fault localization information. Specifically, we employed a perfect fault localization setting in which the actual buggy location is given to NPR techniques. This approach allows us to assess the repair capability of NPR techniques without the bias introduced by a specific fault localization tool, as suggested and employed by prior works~\cite{zhu2021syntax, chen2019sequencer, ye2022neural, xia2022less}.

\vspace{2mm}

\textbf{Patch Correctness Assessment.} 
Following this common practice in prior works~\cite{xia2022less, ye2022neural, zhu2021syntax, lutellier2020coconut}, we employed a two-stage process to evaluate the correctness of patches generated by NPR techniques. Initially, we applied the original patch validation method as implemented in NPR techniques. This validation process continued until a plausible patch, one that successfully passed all test cases, was identified. Subsequently, we conducted a manual assessment of the correctness of this plausible patch. To minimize the potential for errors in this evaluation, each patch is assessed independently by two authors. A patch is deemed correct only if both authors independently confirm its correctness. To ensure the reliability of our assessment, we calculate Cohen's Kappa coefficient, showing almost perfect agreement between two annotators with a score of 0.919. We have published all the patches generated by NPR techniques and our assessment results in our replication package~\cite{package}.

\subsection{Evaluation Metrics on Robustness} 

\subsubsection{Prediction Changes.} Following prior works on robustness evaluation of AI models~\cite{rabin2021generalizability, zhou2022natural}, we use \emph{prediction changes} of models under test as a proxy metric to measure the robustness of NPR techniques. Particularly, we define the prediction changes of an NPR system based on changes in the quality of the NPR-generated patches induced by natural semantic-preserving transformatiosn. The quality of a patch includes the following three categories: (1) \textbf{Wrong Patch} that fails to compile or does not pass the provided test suite; (2) \textbf{Plausible Patch} that successfully passes the given test suite but is deemed \textit{incorrect} in the manual patch correctness assessment; and (3) \textbf{Correct Patch} that both pass the given test suite and is deemed \textit{correct} in the manual patch correctness assessment. We assigned a quality score to each category based on its level of correctness: 0 for a Wrong Patch, 1 for a Plausible Patch, and 2 for a Correct Patch. Given these quality scores, we define the prediction change by an NPR system as follows:
\begin{tcolorbox}
\begin{definition}
    (Prediction Change) A code transformation is considered to induce a prediction change by an NPR system if the quality of the patch for the transformed program is different from the quality of the patch for the original one. 
~\label{def:prediction}
\end{definition}
\end{tcolorbox}

Moreover, in our experiments, we also found that code transformations can affect NPR system predictions in both positive and negative ways. To differentiate their impact on NPR techniques, we introduce two types of changes: \textbf{positive} and \textbf{negative} changes, which denote the prediction changes that offer the higher/lower quality score of generated patches, respectively.

\subsubsection{Performance Changes.} Besides, we also measure the impact of prediction changes on the overall performance of Neural Program Repair systems. Particularly, we calculate the number of bugs that can be repaired correctly and plausibly by NPR techniques for original and transformed bugs. Note that an original bug may have multiple transformed bugs. To calculate the number of correct patches for these transformed bugs, we employ a normalization process that reflects the proportion of fixed bugs out of the total transformed bugs. For instance, if a bug has 5 transformed bugs and the NPR system successfully fixes 3 of them, we consider the number of correct patches to be 0.6 (3/5).

\subsection{Automated Naturalness Assessment of Code Transformations}

\subsubsection{Naturalness Assessment Metrics.} To assess the naturalness of software programs, Hindle et al.~\cite{hindle2016naturalness} proposed to evaluate the predictability and repeatability of their corresponding source regarding a large code corpus. Particularly, the authors measure the naturalness of a code snippet $c$ based on the \textbf{cross-entropy (CE)} metric as
\begin{equation}
    H_{\mathcal{M}}(c) = -\dfrac{1}{n}\sum^{n}_{i=1}log p_{\mathcal{M}}(t_{i}|t_{1}...t_{i-1})
\end{equation}
, where $\mathcal{M}$ is a statistical language model built based on a large corpus of software programs; $t_{1}, t_{2},..., t_{n}$ are tokens of the given code snippet $c$. This metric measures how ``surprised'' the model $\mathcal{M}$ is by the given code snippet~\cite{hindle2016naturalness}. A lower value of \textbf{CE} suggests the less surprise of the model, or in other words the more natural the given code.

\subsubsection{Language Model Selection} Prior works~\cite{hindle2016naturalness, rahman2019natural} typically use \textbf{n-gram} models to assess the naturalness of software code. This model estimates the probability of the nth token given the previous n-1 tokens of a code snippet based on a large corpus of code. In this study, we follow prior works~\cite{hindle2016naturalness, rahman2019natural} to build a n-gram model with modified Kneser-Ney smoothing and $n=4$. We select this value for $n$ because experiments conducted by Hindle et al.~\cite{hindle2016naturalness} and Rahman et al.~\cite{rahman2019natural} have demonstrated that the performance of n-gram models tends to stabilize at this value. Furthermore, increasing n beyond 4 results in limited returns in terms of performance improvement, while significantly increasing memory and time requirements. To train our n-gram models, we use the dataset provided by Rahman et al. which comprises 16 projects with 26,934 Java files. Note that, as our primary focus is at the method level, we divided these files into 297,603 individual methods and trained our n-gram model based on these method-level code snippets.

Besides n-gram models, we also propose to use autoregressive Large Language Models that also estimate a token based on its prefix tokens. As these models are trained on a very large corpus of code snippets, they offer a high-quality estimation for code naturalness through the cross-entropy metric. Particularly, in this work, we consider three well-known and open-sourced autoaggressive Large Language Models (LLM) including GPTNeo~\cite{gptneo}, BLOOM~\cite{scao2022bloom} and CodeLlama~\cite{roziere2023code}. The first two are implemented based on GPT-3  architecture~\cite{brown2020language} and trained on a very large database curated by EleutherAI and  BigScience Workshop, respectively. Meanwhile, the last one is built on top LLama2~\cite{touvron2023llama} and trained on a code-specific dataset containing the most popular programming languages including Java, Python, and C++.
    

\subsubsection{Implementation Details} To conduct experiments on the aforementioned language models, we implement the n-grams model using KenLM~\cite{heafield2011kenlm}, a well-known library for n-gram estimation and Large Language Models using Hugging Face~\footnote{https://huggingface.co/}, a widely-used framework for open-source LLMs. Regarding the version of LLM models, we utilize GPTNeo 2.7B, BLOOM 7B, and CodeLlama 7B parameters for our experiments. All experiments are conducted on 2 x NVIDIA A100 with 40GB of graphic memories and an Intel(R) Xeon(R) Gold 6326 Processor with 100 GB RAM. 
\section{Study Results}
\label{sec:findings}

This section presents the findings from our study, excluding those related to RQ1, which are presented in Section~\ref{sec:criteria}.

\subsection{RQ2: Naturalness of Semantic-preserving Transformations}
~\label{sec:rq1}

\begin{table}[]
    \centering
    \caption{Time taken (in seconds) by annotators for different transformation levels}~\label{tab:time_level}
    \begin{tabular}{l|lll}
        
        \toprule
                        \textbf{Level} & \textbf{Naming} & \textbf{Expression } & \textbf{Statement} \\\midrule
        \textbf{Mean}(s)     & 14.71 & 17.04 & 17.86  \\
        \textbf{Stdev}(s) & 14.78 & 14.51 & 13.77  \\
        \bottomrule
        \end{tabular}
\end{table}
\begin{table}[t]
  \centering
    \caption{Time taken (in seconds) by annotators for high-agreement and disagreement cases}~\label{tab:time_agreement}
  \begin{tabular}{l|ll}
    
    \toprule
                    \textbf{Level} & \textbf{High Agreement}(s) & \textbf{Disagreement}(s) \\\midrule
    \textbf{Mean}  & 15.02 & 16.31  \\
    \textbf{Stdev} & 13.93 & 17.09  \\
    \bottomrule
    \end{tabular}
\end{table}

    
    

\subsubsection{Assessment Quality.}~\label{sec:assess_quality} We ensure the reliability of human annotations in our study by performing sanity checks, following Le et. al~\cite{le2019reliability}. Particularly, we first apply two sanity checks by analyzing completion time to ensure that participants are not arbitrary in their decisions. We expect participants to spend more time assessing more complex transformations. We consider the level of transformations, i.e.,  naming, expression, and statement, as a proxy for transformation complexity. From Table~\ref{tab:time_level}, we can see that participants spent more time on more complex transformations. We also employ the Mann-Whitney-Wilcoxon (MWW) test to validate the statistical significance of this finding. The results indicate their differences are statistically significant with p-values less than 0.05 and effect sizes of 0.22, 0.31, and 0.43 (small to medium) for pairs of levels. Second, we expect participants to spend more time assessing more difficult transformations. We consider the degree of agreement as a proxy for transformation difficulty. Particularly, we compare the completion time of transformations with high agreement, i.e., reaching a consensus of at least 4 out 5 participants, and disagreement, i.e., with only from 2 to 3 participants agreeing with others. From Table~\ref{tab:time_agreement}, it can be seen that participants tend to spend more time on disagreement cases. Their differences are also statistically significant with a p-value of a Mann-Whitney-Wilcoxon (MWW) test of less than 0.05 and an effect size of 0.03 (very small). These results indicate that our participants are not arbitrary in their assessment, indicating the quality of our human study. 

\vspace{2mm}

\begin{table}[t]
    \caption{Agreement levels of participants. "Full Agreement", "High Agreement" and "Disagreement" indicates data points with a consensus of all 5 participants, at least 4 out of 5 participants and only 3 participants, respectively.}~\label{tab:agreement}
    \begin{tabular}{l|lll}
    
    \toprule
                    & \textbf{Full Agreement} & \textbf{High Agreement} & \textbf{Disagreement} \\\midrule
    \textbf{Unnatural} & 42 (3.8\%)  & 214 (19.5\%) & 227 (20.7\%) \\
    \textbf{Natural} & 320 (29.1\%) & 641 (58.4\%) & 16 (1.5\%) \\
    \textbf{Total}  & 362 (32.9\%) & 855 (77.7\%) & 243 (22.3\%) \\
    \bottomrule
    \end{tabular}
\end{table}

\subsubsection{Inter-rater Agreement.}~\label{sec:assess_agreement} Next, we aim to measure the degree of agreement of participants in assessing the naturalness of code transformations.
To recap, our study engaged 5 independent professional developers to evaluate each code transformation. We define a code transformation as receiving a high agreement when at least four out of five participants reach a consensus, while the rest are considered disagreements. 
Table~\ref{tab:agreement} summarizes the number of agreements and disagreements among participants in our human study. 
We can see that participants tend to reach a consensus in the majority of cases (78.1\%) for code transformations, with one-third of them achieving full agreement among all five participants. However, we also observed that disagreement between human annotators, i.e., only 2 or 3 participants agree with others, in approximately one-fifth of code transformations. In addition, these assessments obtained an inter-rater agreement of 0.32~\cite{fleiss1971measuring} calculated by Fleiss's Kappa, an extension of Cohen's Kappa \cite{cohen1960coefficient} for multi-raters. While this score indicates a fair agreement between human annotators, we believe that this degree of agreement is acceptable, given the sensitive nature of naturalness assessment, in which a subset of code transformations presents more challenges for human annotators than others. 

\subsubsection{Naturalness Categories}
To ease our presentation in follow-up sections, we define the naturalness categories for code transformations based on the assessment of human participants as follows:
\begin{itemize}
    \item \textbf{Natural} transformations indicate ones that are deemed as natural by at least 4 out of 5 participants. 
    \item \textbf{Unnatural} transformations indicate ones that are deemed as unnatural by at least 4 out of 5 participants. 
    \item \textbf{Likely Natural} transformations indicate ones that are deemed as natural by 3 out of 5 participants. 
    \item \textbf{Likely Unnatural} transformations indicate ones that are deemed as unnatural by 3 out of 5 participants. 
\end{itemize}
To mitigate potential bias from data points with disagreement, we consider the outcome of their naturalness assessment as ``unknown'' and categorize them as ``likely unnatural'' and ``likely natural'' depending on the majority of assessment. This categorization allows us to distinguish them from cases with high agreement in subsequent experiments.

\subsubsection{Naturalness Assessment.} In this experiment, we investigate the naturalness of semantic-preserving transformations based on the assessment provided in human study. Table~\ref{fig:naturalness_proportion} illustrates the proportion of naturalness categories including unnatural, natural, and unknown in different transformation levels. The 'unknown' category indicates likely unnatural/natural transformations that received disagreement from human annotators.  

\begin{table}[t]
\caption{Proportion of naturalness categories in different transformation levels. \textbf{Unnatural} and \textbf{Natural} indicate likely unnatural/natural transformations received high agreement (consensus of at least 4/5 annotators). \textbf{Unknown} indicate likely unnatural/natural transformations received disagreement from human annotators}~\label{fig:naturalness_proportion}
\begin{tabular}{l|lll|l}
\toprule
          & \textbf{Naming} & \textbf{Expression} & \textbf{Statement} & \textbf{Total} \\ \midrule
\textbf{Unnatural} & 136 (15.9\%)      & 10 (7.1\%)    &  68 (65.4\%)       & 214 (19.5\%)  \\
\textbf{Natural}   & 521 (61\%)      & 94 (67.1\%)     & 26 (25\%)         & 641 (58.4\%)  \\
\textbf{Unknown}   & 197 (23.1\%)    & 36 (25.8\%)    & 10 (9.6\%)        & 243 (22.2\%) \\ \bottomrule
\end{tabular}
\end{table}

Overall, we can see that only 58.8\% of semantic-preserving transformations are deemed natural by human annotators. Additionally, 19.3\% of these transformations are considered unnatural while the remaining 21.9\% is categorized as likely natural/unnatural as they receive a disagreement among human annotators. 

\begin{lstlisting}[float, language=java, caption=A example of an unnatural bug created by a ReverseIf transformation which reverse then-block and else-block an if-statement on the bug Mockito-22 in Defects4J dataset, label=lst:statement_example]
//Simplified Original Code
    if (o1 == null || o2 == null) {
      return o1 == null && o2 == null;
    } else if (isArray(o1)) {
      return isArray(o2) && areArraysEqual(o1, o2);
    } else {
      return o1.equals(o2);
    }
//Simplified Transformed Code
    if !((o1 == null || o2 == null)) {
      if (isArray(o1)) {
        return isArray(o2) && areArraysEqual(o1, o2);
      } else {
        return o1.equals(o2);
      }
    } else {
      return o1 == null && o2 == null;
    }
\end{lstlisting}
In an in-depth analysis of each level of transformation, we found that statement-level transformations exhibit the highest proportion of unnatural transformations, accounting for 62.4\% of total transformations, while natural transformations occupy only 27.4\% of this category. This is because these transformations usually change the code structure and require complex transformations to preserve the original code semantics. For example, Listing~\ref{lst:statement_example} illustrates an unnatural \texttt{ReverseIf} transformation. 
This transformation reverses the contents of both the then-block and the else-block within an if-statement from the original code, resulting in a newly generated code that preserves the original code's semantics. In this example, the transformation relocates the else-if-else-block found in lines 6-10 of the original code to inside the then-block of the new code, spanning lines 14-18. While this transformation successfully preserves the code semantics, it does disrupt the structure of the original code. Consequently, 4 out of 5 annotators agreed that it reduces the readability of the original code, rendering it unnatural. 

\begin{lstlisting}[float, language=java, caption=A example of an unnatural program created by a variable renaming transformation which changes variable name v to V on the bug Lang-22 in Defects4J dataset. The bug in this program happens in line 3 (and the corresponding line 10 in the transformed program) in which developers originally assume that u and v are non-zero numbers so they do not handle the cases of zero values and cause logical errors., escapechar=!, label=lst:variable_example]
//Simplified Original Code
  private static int greatestCommonDivisor(int u, int !\hl{v}!) {
    if (Math.abs(u) <= 1 || Math.abs(!\hl{v}!) <= 1) {
      return 1;
    }
    ...
  }
//Simplified Transformed Code
  private static int greatestCommonDivisor(int u, int !\sethlcolor{mGreen} \hl{V}!) {
    if (Math.abs(u) <= 1 || Math.abs(!\sethlcolor{mGreen}\hl{V}!) <= 1) {
      return 1;
    }
    ...
  }
\end{lstlisting}

Regarding the two remaining transformation levels, i.e., naming and expression, naming-level transformations exhibit a higher ratio of unnatural code transformations compared to expression-level transformations. This arises from the fact that, while naming-level transformations are less complex, they necessitate the identification of suitable substitutions for variable names. Though these substitutions are studied in terms of naturalness by prior works~\cite{zhou2022natural, gerobustnpr}, it is not perfect and may still produce unnatural substitutions. For instance, Listing~\ref{lst:variable_example} illustrates a substitution generated by the \texttt{Variable-Renaming-2} transformation proposed by Zhou et al.~\cite{zhou2022natural}, which leverages the masked token prediction feature of CodeBERT~\cite{zhangyin2020codebert} to produce suitable substitutions. Unfortunately, in this particular case, CodeBERT generates an unnatural substitution that breaks the common coding convention, rendering this transformation unnatural.

\find{\textbf{Finding 1:} There are only 58.8\% of semantic-preserving transformations that are natural regarding human assessment while the other 19.3\% are deemed unnatural. Additionally, among transformation levels, statement-level transformation exhibits the highest unnatural ratio with 62.4\% of transformation at this level are consider unnatural by human assessment.}

\subsection{RQ3: Impact of Unnaturalness in Semantic-Preserving Transformations on Robustness Testing} \label{sec:rq2}

In this experiment, we aim to assess the effect of unnaturalness in semantic-preserving transformations on robustness testing. \thanh{To achieve this, we first evaluate how unnaturalness influences the applicability of semantic-preserving transformations on our dataset (Section~\ref{sec:rq3_coverage}). Next, we examine its effect on prediction changes  (Section~\ref{sec:rq3_prediction_changes}) and performance changes (Section~\ref{sec:rq3_performance_changes}) of NPR techniques.}

\begin{figure}
    \centering
    \caption{Applicability of semantic-preserving transformations on 220 bugs from the Defects4J dataset. \textbf{Naming}, \textbf{Expression}, and \textbf{Statement} represent the number of applicable bugs for their respective categories of code transformation, while \textbf{All} presents results from all categories.}
    \label{fig:applicability}
    \includegraphics[width=0.8\linewidth]{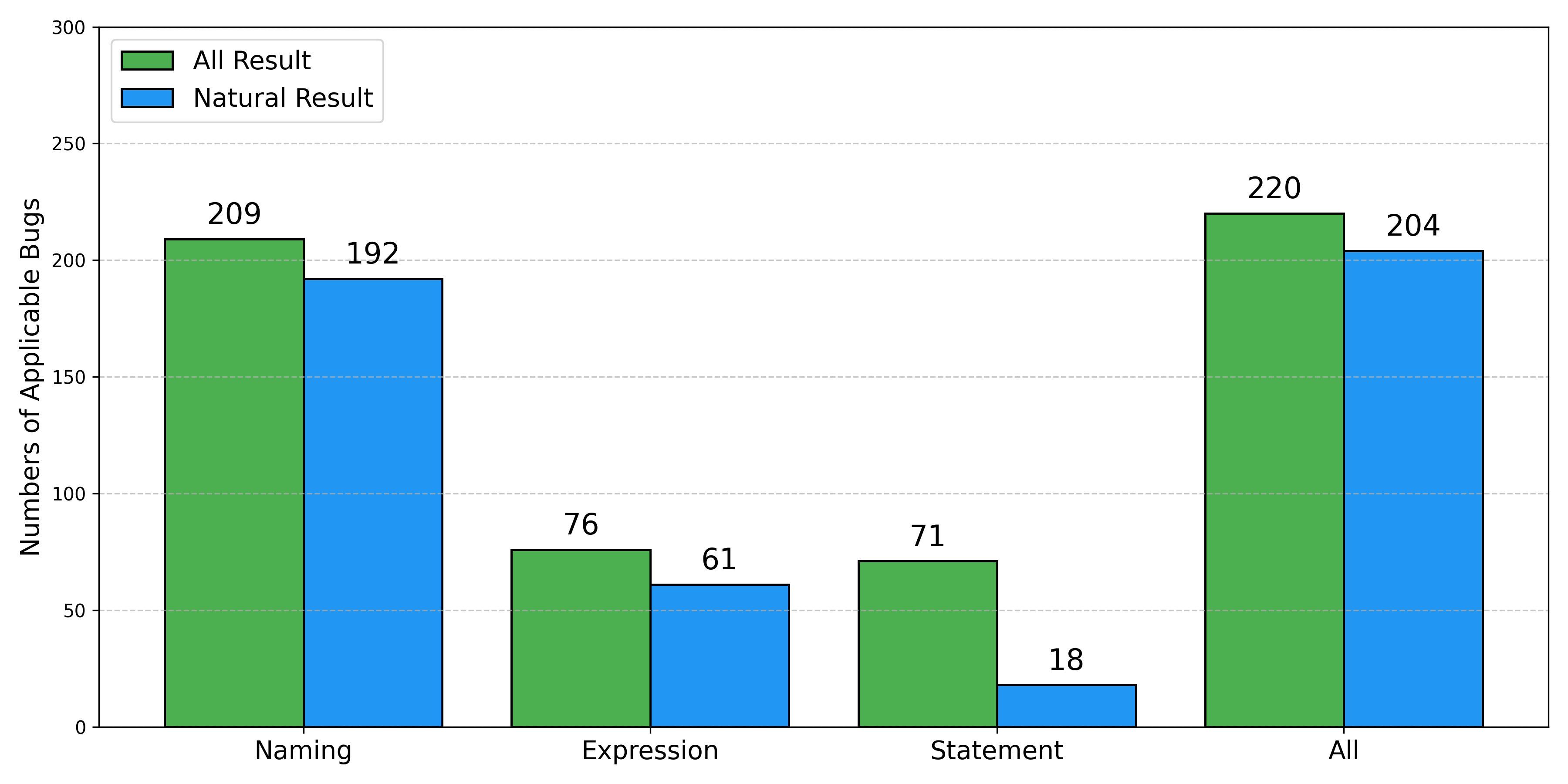}
\end{figure}

\vspace{1mm}

\subsubsection{Applicability}~\label{sec:rq3_coverage}
\thanh{Figure~\ref{fig:applicability} illustrates the impact of naturalness on the applicability of semantic-preserving transformations in our datasets. The overall impact of unnaturalness on the applicability of these transformations is moderate, with a slight reduction from 220 applicable bugs when using all semantic-preserving transformations to 204 applicable bugs when restricting to only natural transformations. However, the effect of unnaturalness varies across different transformation categories. In the \textbf{naming} and \textbf{expression} categories, the difference between the two approaches is relatively small (209 vs. 192 and 76 vs. 61 applicable bugs). However, in the \textbf{statement} category, natural transformations exhibit a more pronounced drop in applicability, falling from 71 applicable bugs to only 18. This is expected, as previous experiments revealed that 65.4\% of statement-level transformations tend to be unnatural. This finding highlights a lack of naturalness in the design of statement-level transformations, which often involve more complex modifications. It underscores the importance of incorporating naturalness considerations when designing such transformations to enhance their applicability for natural robustness testing.}

\begin{figure}
    \centering
    \caption{Proportion of naturalness categories in prediction changes of NPR techniques. \textbf{Unnatural} and \textbf{Natural} indicate likely unnatural/natural transformations received high agreement (consensus at least of 4/5 annotators). \textbf{Likely Natural/Unnatural} indicate likely unnatural/natural transformations received disagreement from human annotators}
    \label{fig:naturalness_proportion_on_change}
    \includegraphics[width=0.9\linewidth]{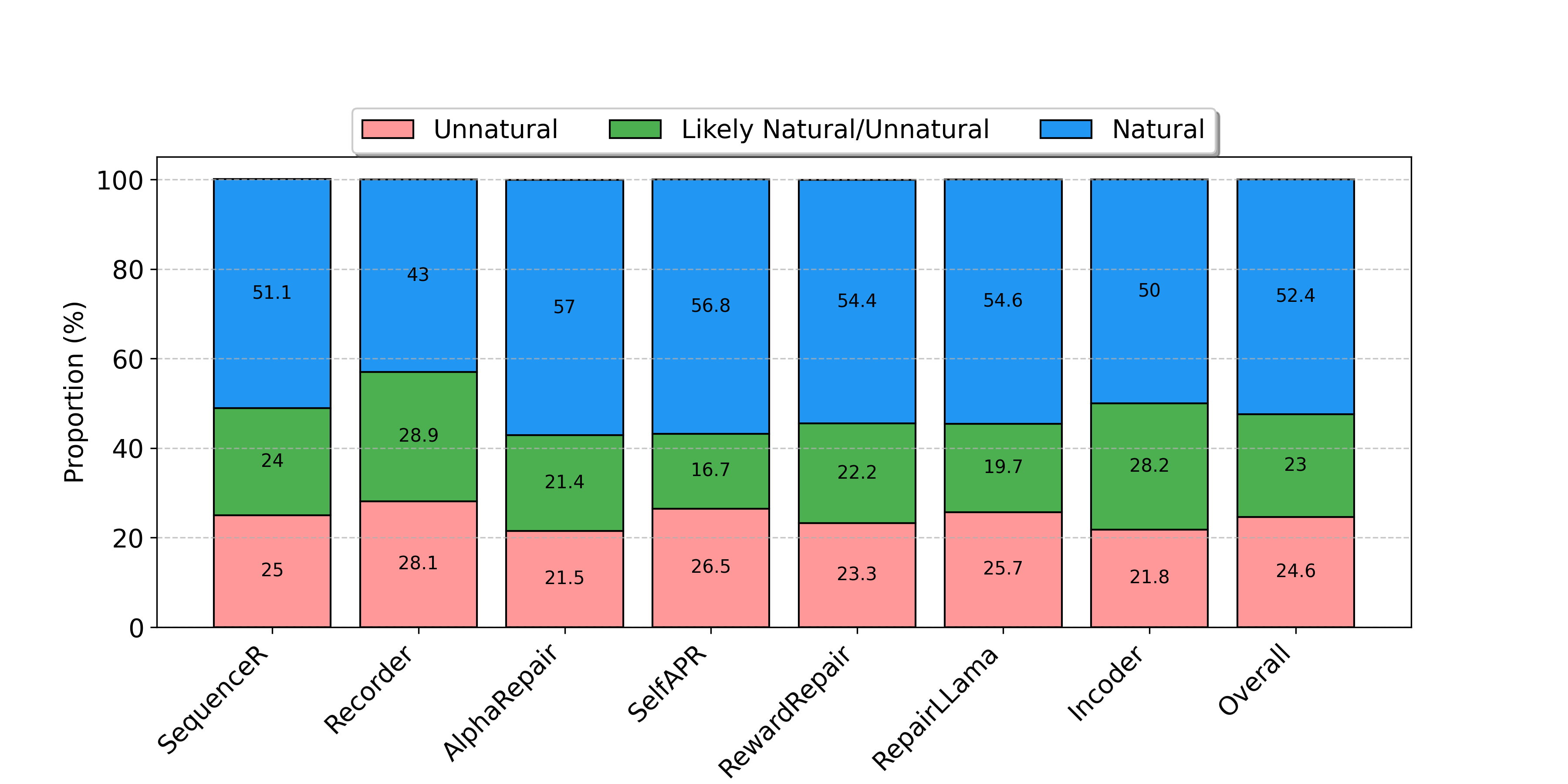}
\end{figure}

\find{\textbf{Finding 2:} Unnaturalness in the design of semantic-preserving transformations significantly affects the applicability of statement-level transformations in natural robustness testing. However, its overall impact is moderate, as naming and expression transformations are only slightly affected.
}

\vspace{1mm}

\subsubsection{Prediction Changes.}~\label{sec:rq3_prediction_changes} Next, we examined the impact of unnaturalness in semantic-preserving transformations on prediction changes in NPR techniques. Specifically, we identified prediction changes in NPR techniques following Definition~\ref{def:prediction}, wherein a code transformation is considered to induce a prediction change if the quality of the generated patch for the transformed program differs (either improves or degrades) from the patch generated for the original program. We then categorized these prediction changes based on their naturalness, as assessed through a human evaluation process described earlier (Section~\ref{sec:rq2}).

\begin{figure}[t]
    \centering
    \caption{Prediction Changes of NPR techniques against all transformation (denoted by All Results) and natural transformations (denoted by Natural Results)}
    \label{fig:prediction_changes_compare}
    \begin{subfigure}{0.24\textwidth}
        \centering
        \includegraphics[width=\textwidth]{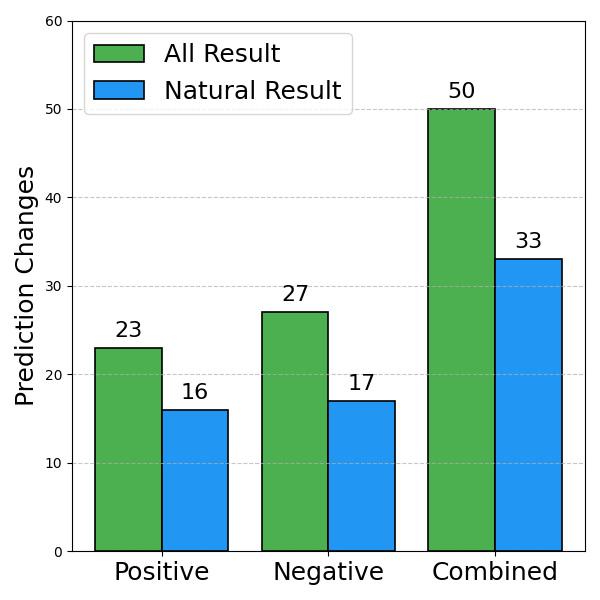}
        \caption{AlphaRepair}
        \label{fig:sub1}
    \end{subfigure}
    \hfill
    \begin{subfigure}{0.24\textwidth}
        \centering
        \includegraphics[width=\textwidth]{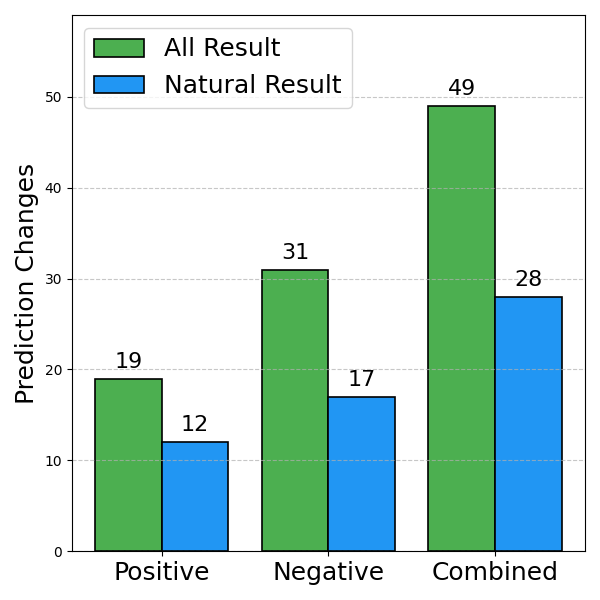}
        \caption{Recoder}
        \label{fig:sub2}
    \end{subfigure}
    \hfill
    \begin{subfigure}{0.24\textwidth}
        \centering
        \includegraphics[width=\textwidth]{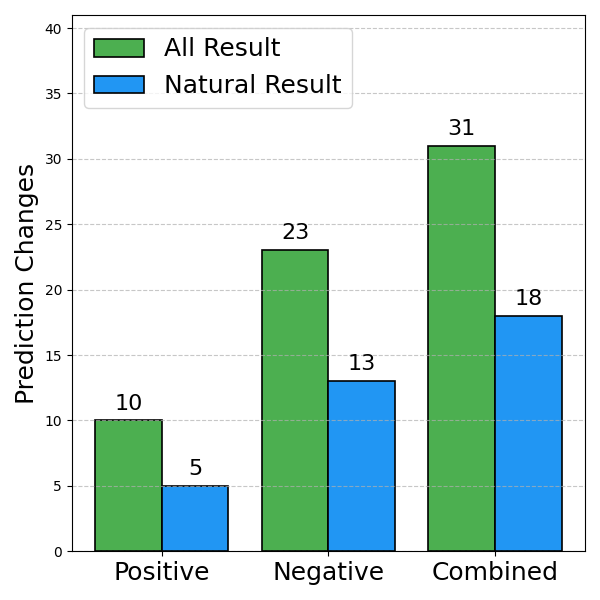}
        \caption{SequenceR}
        \label{fig:sub3}
    \end{subfigure}
    \hfill
    \begin{subfigure}{0.24\textwidth}
        \centering
        \includegraphics[width=\textwidth]{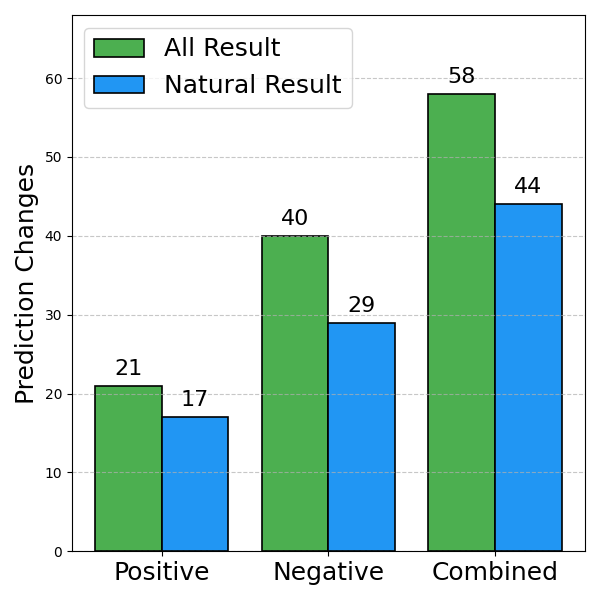}
        \caption{SelfAPR}
        \label{fig:sub4}
    \end{subfigure}

    \vspace{0.5cm}
    
    \begin{subfigure}{0.24\textwidth}
        \centering
        \includegraphics[width=\textwidth]{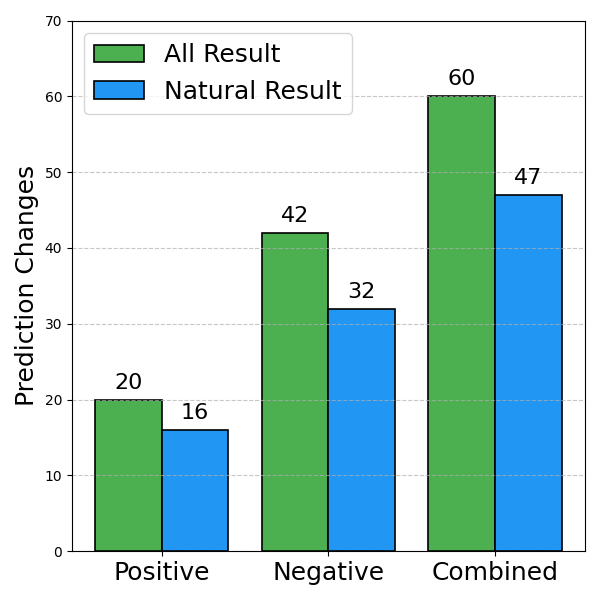}
        \caption{RewardRepair}
        \label{fig:sub5}
    \end{subfigure}
    \hfill
    \begin{subfigure}{0.24\textwidth}
        \centering
        \includegraphics[width=\textwidth]{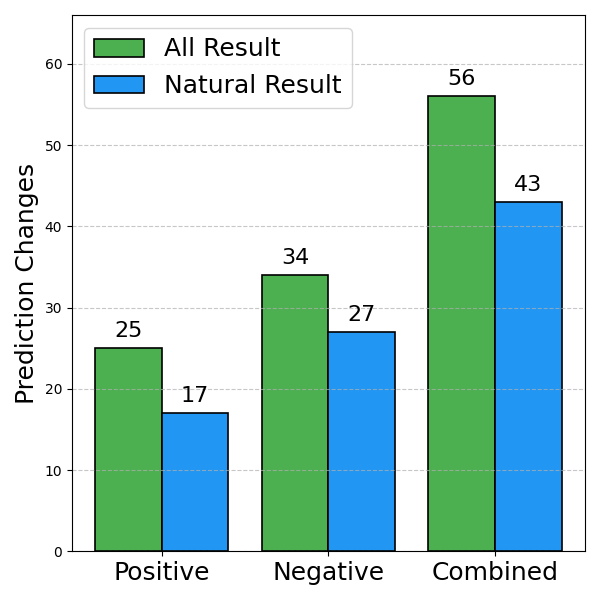}
        \caption{Incoder}
        \label{fig:sub6}
    \end{subfigure}
    \hfill
    \begin{subfigure}{0.24\textwidth}
        \centering
        \includegraphics[width=\textwidth]{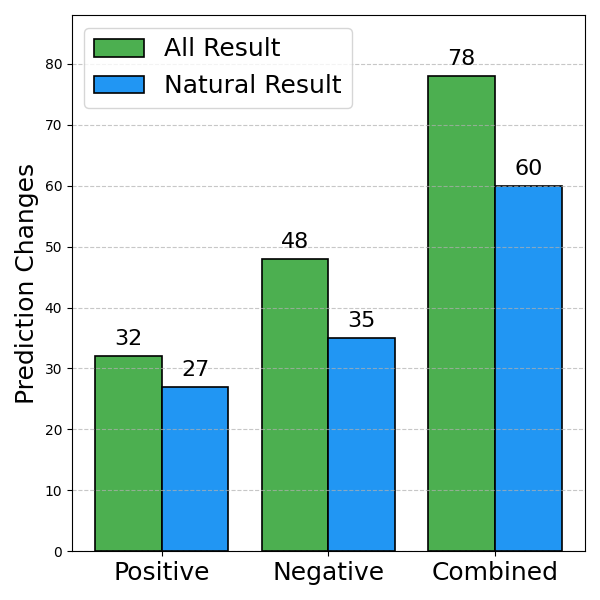}
        \caption{RepairLLama}
        \label{fig:sub7}
    \end{subfigure}
\end{figure}

The results, as illustrated in Figure~\ref{fig:naturalness_proportion_on_change}, reveal that unnatural transformations account for 24.6\% of prediction changes across all NPR techniques. This suggests that approximately a quarter of prediction changes are linked to transformations that deviate from what humans perceive as natural code—transformations and rarely occurs in real-world projects. Consequently, focusing on improving these unnatural transformations may lead to wasted effort, as their practical impact is minimal. In this work, we refer to such instances as false alarms from robustness testing. The false alarm rate due to unnatural transformations varies across different NPR techniques, ranging from 21.5\% (for SelfAPR) to 28.1\% (for Recorder).

Furthermore, considering our goal of leveraging natural robustness testing for more reliable NPR evaluation, we might expect that most instances of NPR unrobustness (i.e., prediction changes) would arise from natural transformations—code changes that align with human expectations of naturalness. However, our findings show that only 52.4\% of prediction changes result from natural transformations. This observation also hold for all NPR techniques with proportion of prediction changes caused by natural code transformation ranging from 43\% to 56\%.

\thanh{Additionally, we also further investigate the differences in findings of robustness testing with all transformations and only natural transformations, as illustrated in Figure~\ref{fig:prediction_changes_compare}. Our analysis shows that robustness testing with all transformations (referred to as complete robustness) tends to yield greater prediction changes across various Neural Program Repair (NPR) techniques compared to testing with only natural transformations. For example, in AlphaRepair, the positive and negative prediction changes observed with complete robustness are 43\% and 58\% higher, respectively, than those recorded with natural robustness. This pattern is consistently observed across all NPR techniques. These results indicate that complete robustness testing with all semantic-preserving transformations may produce misleading signals regarding the actual performance of NPR techniques.}

\thanh{These findings underscore the importance of accounting for transformation naturalness when evaluating the robustness of NPR techniques. Differentiating whether robustness issues stem from unnatural or natural transformations could enable more precise assessments of NPR performance.}

\find{\textbf{Finding 3:} Unnatural code transformations result in a false alarm rate of 24.6\% during the robustness evaluation of NPR techniques. Additionally, only 52.4\% of instances of unrobustness are due to natural transformations. Consequently, robustness testing using all semantic-preserving transformations leads to higher prediction changes than those observed in natural robustness testing, introducing misleading signals about the actual performance of NPR techniques.
}

\begin{table}[t]
    \centering
    \caption{Performance changes on plausible and correct rate compared to original evaluation dataset in robustness testing with all transformations and only natural transformations.}
    \label{tab:performance_change_compare}
    \begin{tabular}{|l|l|l|l|l|}
        \toprule
        \textbf{Repair Tool} & \multicolumn{2}{c|}{\textbf{Plausible Rate}} & \multicolumn{2}{c|}{\textbf{Correct Rate}} \\ \hline
        
        & \textbf{All} & \textbf{Natural} & \textbf{All} & \textbf{Natural} \\ \midrule
        
        \texttt{AlphaRepair}   & -7.7\% & -2.9\%  & -17.9\% & -11.8\% \\
        
        \texttt{Recoder}       & -10.4\% & -4.3\% & -14.0\% & -6.5\% \\
        
        \texttt{SequenceR}     & -19.0\% & -17.0\% & -31.4\% & -29.7\% \\
        
        \texttt{SelfAPR}       & -24.3\% & -14.6\% & -25.5\% & -22.5\% \\
        
        \texttt{RewardRepair}   & -15.8\% & -12.4\% & -35.6\% & -28.6\% \\
        
        \texttt{Incoder}       & -10.8\% & -6.5\% & -9.0\% & -1.5\% \\
        
        \texttt{RepairLLama}   & -9.8\% & -6.6\% & -23.2\% & -23.1\% \\ \bottomrule
        
    \end{tabular}
\end{table}

\subsubsection{Performance Changes.} \thanh{Finally, we examine the impact of unnaturalness on the performance variations of Neural Program Repair (NPR) techniques, specifically focusing on the relative differences in their performance when evaluated on transformed datasets compared to the original dataset. In particular, we measure the performance changes in terms of plausible and correct rates during robustness testing, considering all transformations as well as only natural transformations, as illustrated in Table~\ref{tab:performance_change_compare}.}

\thanh{We observe that the performance changes of NPR techniques when evaluated against all transformations are significantly larger than those observed with only natural transformations. In terms of plausible rates, NPR techniques exhibit drops ranging from 7.7\% to 24.3\% under all transformations, while the declines are notably smaller, ranging from 2.9\% to 17\% when limited to natural transformations. A similar trend is also observed in the correct rates, where performance changes under all transformations range from 9.0\% to 35.6\%, compared to only 1.5\% to 29.7\% for natural transformations.}

\thanh{Furthermore, we observe that the comparison of performance changes between NPR techniques differ significantly when evaluated under all transformations compared to only natural transformations. For instance, the performance change in plausible rate for SelfAPR is notably higher than that of RewardRepair under all transformations (24.3\% vs. 15.8\%); however, when considering only natural transformations, their performance changes are nearly equivalent (14.6\% vs. 12.4\%). In contrast, the performance change in plausible rate for SequenceR is 8.6\% higher than that of Recoder under all transformations, but this difference increases to 12.7\% when limited to natural transformations. Similar trends are also observed in many other pairs of NPR techniques.}

\find{\textbf{Finding 4:} Performance changes measured under all transformations are significantly larger than those measured with only natural transformations. Additionally, the comparison of performance changes among NPR techniques varies considerably when evaluated under all transformations versus only natural transformations.
}

\subsection{RQ4: Natural Robustness of NPR techniques.}~\label{sec:rq4}

In this experiment, we aim to investigate the natural robustness of NPR techniques against the 641 natural code transformations identified in our human study. To this end, we measure the proportion of bugs in the Defects4J dataset where prediction changes occur. \thanh{Additionally, we analyze prediction changes across different transformation categories to provide deeper insights into the natural robustness of NPR techniques.} Finally, we assess the impact of natural code transformations on the overall performance of NPR techniques, specifically in terms of the number of plausible and correct patches. 

\begin{table}[t]
\centering
\caption{Prediction Changes of NPR techniques under Natural Code Transformations. \textbf{Positive} and \textbf{Negative} denote the number of bugs with positive and negative prediction changes, respectively. \textbf{All} represents bugs with either positive or negative changes.}
\label{tab:prediction_changes}
\vfill
\begin{tabular}{l|l|l|l}
\toprule
\textbf{Repair Tool} & \textbf{Positive} & \textbf{Negative} & \textbf{All} \\ \midrule
\texttt{AlphaRepair}  & 16 (7.8\%)   & 17 (8.3\%)   & 33 (16.2\%)  \\
\texttt{Recoder}      & 12 (5.9\%)   & 17 (8.3\%)   & 28 (13.7\%)  \\ 
\texttt{SequenceR}    & 5 (2.5\%)    & 13 (6.4\%)   & 18 (8.8\%)   \\ 
\texttt{SelfAPR}      & 17 (8.3\%)   & 29 (14.2\%)  & 44 (21.6\%)  \\
\texttt{RewardRepair} & 16 (7.8\%)   & 32 (15.7\%)  & 47 (23.0\%)  \\ 
\texttt{Incoder}      & 17 (8.3\%)   & 27 (13.2\%)  & 43 (21.1\%)  \\ 
\texttt{RepairLLama}  & 27 (13.2\%)  & 35 (17.2\%)  & 60 (29.4\%)  \\ 
\bottomrule
\end{tabular}
\end{table}

\subsubsection{Prediction Changes.} Table~\ref{tab:prediction_changes} presents a detailed breakdown of the number and proportion of bugs from the Defects4J dataset for which prediction changes occur under natural code transformations, encompassing both positive and negative changes.

As observed, all NPR (Neural Program Repair) techniques exhibit some level of prediction variation when subjected to these transformations. The number of bugs affected by prediction changes ranges from 18 for SequenceR to 60 for RepairLLama, representing between 8.8\% and 29.4\% of the total 204 bugs in the Defects4J dataset. This wide range of impact highlights the lack of natural robustness across different repair techniques. \thanh{Notably, LLM-based models such as Incoder and RepairLLama, despite their extensive pre-training on large-scale corpora, are still affected by natural code transformations with 21.1\% and 29.4\% bugs witnessed prediction changes. This can likely be attributed to their dependence on smaller, domain-specific datasets for fine-tuning or in-context learning during program repair tasks. While pre-training enables these models to capture broad patterns across diverse programming contexts, the limited size and scope of the repair datasets may not fully prepare them for handling subtle, real-world variations in code. Consequently, these models remain vulnerable to the impact of natural code transformations, which can degrade their natural robustness.}

A key observation from the table is that natural code transformations do not solely induce negative changes in predictions. In fact, all techniques show a proportion of positive prediction changes, meaning that the quality of their output actually improves after the transformation. For instance, positive changes range from 10 bugs (2.5\%) for SequenceR to 32 bugs (13.2\%) for RepairLLama. This demonstrates that these transformations sometimes help NPR models make better predictions, possibly by introducing variations that align better with the underlying model's learned patterns.

However, despite the presence of positive changes, the majority of prediction changes remain negative. Negative prediction changes occur in a larger proportion of bugs, ranging from 6.4\% (for SequenceR) to 17.2\% (for RepairLLama). This indicates that while some transformations can enhance the output, many tend to degrade the model's predictions, emphasizing the importance of robustness in these systems when dealing with naturally occurring variations in the code. We will further discuss about the impact of natural robustness on performance of NPR techniques in Section~\ref{sec:impact_on_performance}. 

\find{\textbf{Finding 5:} NPR techniques exhibit a lack of robustness against natural code transformations, resulting in prediction changes for 8.8\% to 29.4\% of target bugs. However, these changes do not exclusively lead to negative outcomes; in some cases, they also improve quality of their predictions.
}

\begin{table}[t]
\centering
\caption{Prediction Changes of NPR techniques across different code transformation categories.}
\label{tab:prediction_changes_category}
\begin{tabular}{l|l|l|l}
\toprule
\textbf{Repair Tool} & \textbf{Naming} & \textbf{Expression} & \textbf{Statement} \\
& \textit{(192 bugs)} & \textit{(61 bugs)} & \textit{(18 bugs)} \\
\midrule
\texttt{AlphaRepair}   & 25 (13.0\%)  & 14 (23.0\%)  & 4 (22.2\%)  \\ 
\texttt{Recoder}       & 20 (10.4\%)  & 11 (18.0\%)  & 3 (16.7\%)  \\ 
\texttt{Sequencer}     & 14 (7.3\%)   & 6 (9.8\%)    & 2 (11.1\%)  \\ 
\texttt{SelfAPR}       & 33 (17.2\%)  & 13 (21.3\%)  & 3 (16.7\%)  \\ 
\texttt{RewardRepair}  & 36 (18.8\%)  & 15 (24.6\%)  & 7 (38.9\%)  \\
\texttt{Incoder}       & 29 (15.1\%)  & 22 (36.1\%)  & 6 (33.3\%)  \\
\texttt{RepairLLama}   & 43 (22.4\%)  & 23 (37.7\%)  & 8 (44.4\%)   \\ 
\bottomrule
\end{tabular}
\end{table}

\subsubsection{Prediction Changes per transformation category.} \thanh{Next, we further analyze the prediction changes of NPR techniques across different categories of code transformations to better understand the impact of each category on the natural robustness of these techniques. Table~\ref{tab:prediction_changes_category} presents our experimental results, highlighting how each tool performs across three transformation categories: Naming, Expression, and Statement.}

\thanh{\textbf{Naming} transformations appear to have a moderate impact on the robustness of NPR techniques, with prediction changes ranging from 7.3\% to 22.4\%. However, despite the lower percentage of prediction changes compared to other categories, the overall impact of naming transformations is the highest across all NPR techniques, except for Sequencer. This is primarily due to the broader coverage of naming transformations, which can be applied to 192 bugs, significantly more than the 61 and 18 applicable bugs for expression and statement transformations, respectively. This wider applicability enlarge the influence of naming transformations on the overall robustness of NPR techniques.}       

\thanh{In contrast, \textbf{expression} transformations result in higher percentages of prediction changes on the applicable bugs compared to naming transformations. Prediction changes in this category range from 9.8\% to 37.7\%, with most tools experiencing at least a 20\% change rate. Meanwhile, \textbf{statement} transformations show the highest average percentages of prediction changes, ranging from 11.1\% to 44.4\%. This is expected, as statement and expression transformations typically affect the overall flow or structure of the code, leading to more substantial changes in bug patterns. Such structural modifications introduce greater challenges for NPR techniques to handle effectively. However, the requirements for transforming code at the statement and expression levels make these transformations less applicable to our dataset, with only 61 and 18 bugs, respectively, compared to 192 bugs for naming transformations. As a result, the overall impact of expression and statement transformations on NPR techniques is lower than that of naming transformations, affecting predictions for less than 23 and 8 bugs, respectively.}

\find{\textbf{Finding 6:} 
Our analysis shows that naming transformations have the highest overall impact on NPR techniques due to their broader applicability, affecting 192 bugs. In contrast, expression and statement transformations tend to be more challenging to NPR techniques but apply to far fewer bugs, limiting their overall impact on natural robustness.}


\begin{table}[t]
    \centering
    \caption{Plausible Rate and Correct Rate of Neural Program Repair techniques on original dataset (denoted by \textbf{Origin}) and transformed dataset with natural code transformations (denoted by \textbf{Transform}).}
    \label{tab:repair_tools_performance}
    \begin{tabular}{l|c|c|c|c|c|c}
        \toprule
        \textbf{Repair Tool} & \multicolumn{3}{c|}{\textbf{Plausible Rate}} & \multicolumn{3}{c}{\textbf{Correct Rate}} \\ \hline
        & \textbf{Origin} & \textbf{Transform} & \textbf{Changes} & \textbf{Origin} & \textbf{Transform} & \textbf{Changes} \\ \hline
        
        \texttt{AlphaRepair}   & 29.9\%                   & 29.1\%                  & -2.7\%     & 13.7\%                 & 12.3\%                & -10.2\%    \\
        
        \texttt{Recoder}       & 24.0\%                   & 23.0\%                  & -4.2\%     & 15.7\%                 & 14.7\%                & -6.4\%     \\
        
        \texttt{SequenceR}     & 19.1\%                   & 16.3\%                  & -14.7\%    & 9.8\%                  & 7.6\%                 & -22.4\%    \\
        
        \texttt{SelfAPR}       & 26.5\%                   & 23.1\%                  & -12.8\%    & 14.7\%                 & 12.0\%                & -18.4\%    \\
        
        \texttt{RewardRepair}  & 33.3\%                   & 29.7\%                  & -10.8\%    & 19.6\%                 & 15.2\%                & -22.4\%    \\
        
        \texttt{Incoder}       & 28.4\%                   & 26.7\%                  & -6.0\%     & 16.2\%                 & 15.9\%                & -1.9\%     \\
        
        \texttt{RepairLLama}   & 42.6\%                   & 40.0\%                  & -6.1\%     & 33.3\%                 & 27.1\%                & -18.6\%    \\ \bottomrule
        
    \end{tabular}
\end{table}

\begin{figure}
    \centering
    \includegraphics[width=\textwidth]{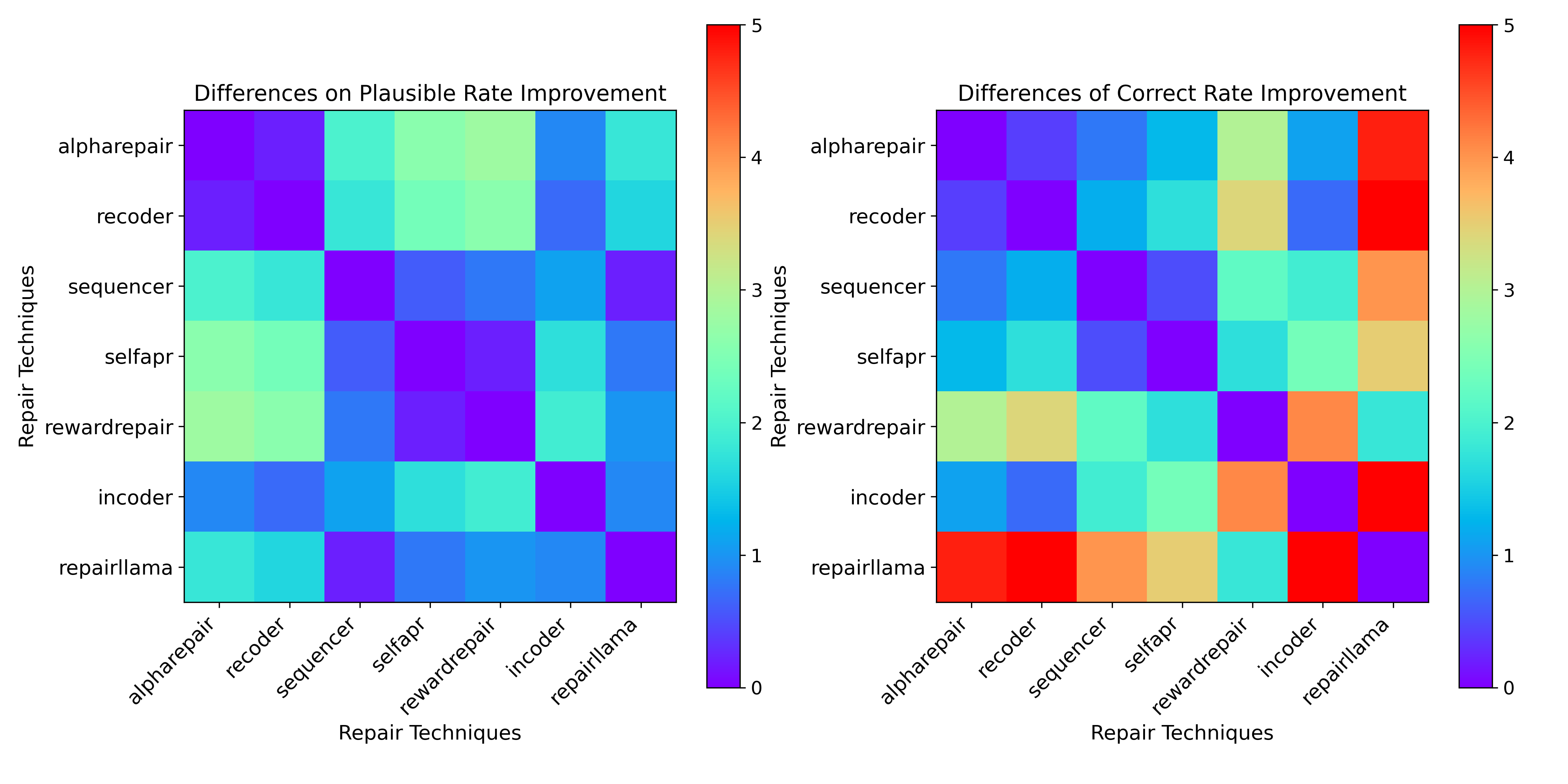}
    \caption{Absolute Differences in Plausible Rate and Correct Rate Improvements between NPR Techniques on Original and Transformed Benchmarks}
    \label{fig:improvement_diffs}
\end{figure}

\vspace{1mm}

\subsubsection{Impact on NPR Performance.} ~\label{sec:impact_on_performance} Finally, we also evaluate the impact of prediction changes identified in previous experiments on the overall performance of Neural Program Repair (NPR) techniques. This is done by measuring two key metrics: the plausible rate and the correct rate. The plausible rate represents the ratio of bugs that can be plausibly fixed by NPR techniques in our dataset, while the correct rate indicates the ratio of bugs that can be correctly fixed. Table~\ref{tab:repair_tools_performance} presents the plausible and correct rates of NPR techniques on both the original dataset and the transformed dataset created by natural code transformations.

The performance of NPR techniques shows a notable decline on the transformed dataset compared to their performance on the original dataset. Specifically, all repair tools exhibit a decrease in plausible rates upon transformation, with reductions ranging from 2.7\% to 14.7\%. For instance, SequenceR demonstrates the most significant drop in plausible rate, falling from 19.1\% to 16.3\%. This indicates a substantial reduction in its ability to propose plausible fixes after applying natural code transformations. 

The situation is particularly critical when considering the correct rate. Notably, NPR techniques experience a more pronounced decrease in correct rates than in plausible rates across all tools, further emphasizing the impact of the observed prediction changes. Techniques such as RewardRepair and SequenceR show declines of 22.4\%, highlighting their reduced effectiveness in generating correct fixes under natural code transformations. Interestingly, Incoder exhibits the least reduction in correct rate at just 1.9\%. In a deeper analysis, we found that this happens because negative prediction changes of Incoder mostly appears in plausible patches and do not affect correct patches much. 

\thanh{Interestingly, Incoder experiences the least reduction in correct rate at 1.9\%, a result that can be attributed to its negative prediction changes predominantly affecting plausible but incorrect patches. Besides, NPR techniques based on Neural Machine Translation (NMT), including SequenceR, SelfAPR, and RewardRepair, exhibit more substantial drops in performance (10-15\% for Plausible Rate and 18.4-22.4\% for Correct Rate) compared to tree-based (Recoder), LLM-based (RepairLLama \& Incoder), and cloze-based (AlphaRepair) techniques, which show smaller declines, typically under 10\%. This suggests that NMT-based methods may be more sensitive to syntactic and structural changes introduced by natural code transformations. We hypothesize that this sensitivity stems from their token-level representation of source code, which, when combined with limited training data, makes it difficult for these models to generalize to unseen code structures and variations. In contrast, more advanced methods, such as LLM-based and tree-based approaches, mitigate this issue by either leveraging large-scale pre-training (in the case of LLMs) or by representing the source code as an abstract syntax tree (AST), capturing more of the underlying structure and semantics.}

\thanh{Furthermore, while the ranking of NPR techniques remains largely consistent, we observed that the varying degrees of decline in performance lead to shifts in relative improvements. the varying degrees of performance decline result in significant shifts in their relative improvements. To better understand these differences, we examined the absolute improvements in Plausible Rate and Correct Rate across the NPR techniques on both benchmarks, as shown in Figure~\ref{fig:improvement_diffs}. The results reveal notable differences in improvement rates, with some exceeding 5\%. For example, on the original dataset, Incoder produced correct patches 3.4\% less effectively than SelfAPR, but on the transformed dataset, Incoder slightly outperformed SelfAPR by 0.7\%. These variations arise from the differing robustness levels of the NPR techniques. For instance, Incoder experienced only a 1.9\% drop in Correct Rate, whereas SelfAPR saw a much larger decline of 22.4\%.} 

\thanh{The absolute differences in improvement rates (2-5\%) may seem small but are significant in NPR evaluations, where typical improvements range from 5-10\%. This emphasizes the potential for performance biases when evaluating NPR tools using limited datasets. Therefore, we advocate for incorporating natural robustness testing through semantic-preserving transformations to enhance the reliability of NPR evaluations and better understand how different techniques handle realistic code variations.}

\find{\textbf{Finding 7:} The evaluation of Neural Program Repair (NPR) techniques reveals a significant decline in both plausible and correct rates when comparing performance on original and transformed datasets, with reductions ranging from 1.9\% to 22.4\% across various tools. Additionally, while the ranking of NPR techniques mostly remains consistent, significant absolute differences in performance improvements between NPR techniques emerge, highlighting potential biases introduced by limited datasets.}

\subsection{RQ5: Effectiveness of Language Models on Automated Naturalness Assessment of Code Transformations}
\label{sec:rq3}

\subsubsection{Initial Experiments on Cross-Entropy}
\begin{figure}
    \centering
    \caption{Distributions of naturalness metrics on different naturalness categories. \textbf{Unnatural} and \textbf{Natural} indicate likely unnatural/natural transformations received high agreement (consensus at least of 4/5 annotators). \textbf{Likely Unnatural/Natural} transformations received disagreement from human annotators.}
    \begin{subfigure}[b]{0.48\textwidth}
        \includegraphics[width=\textwidth]{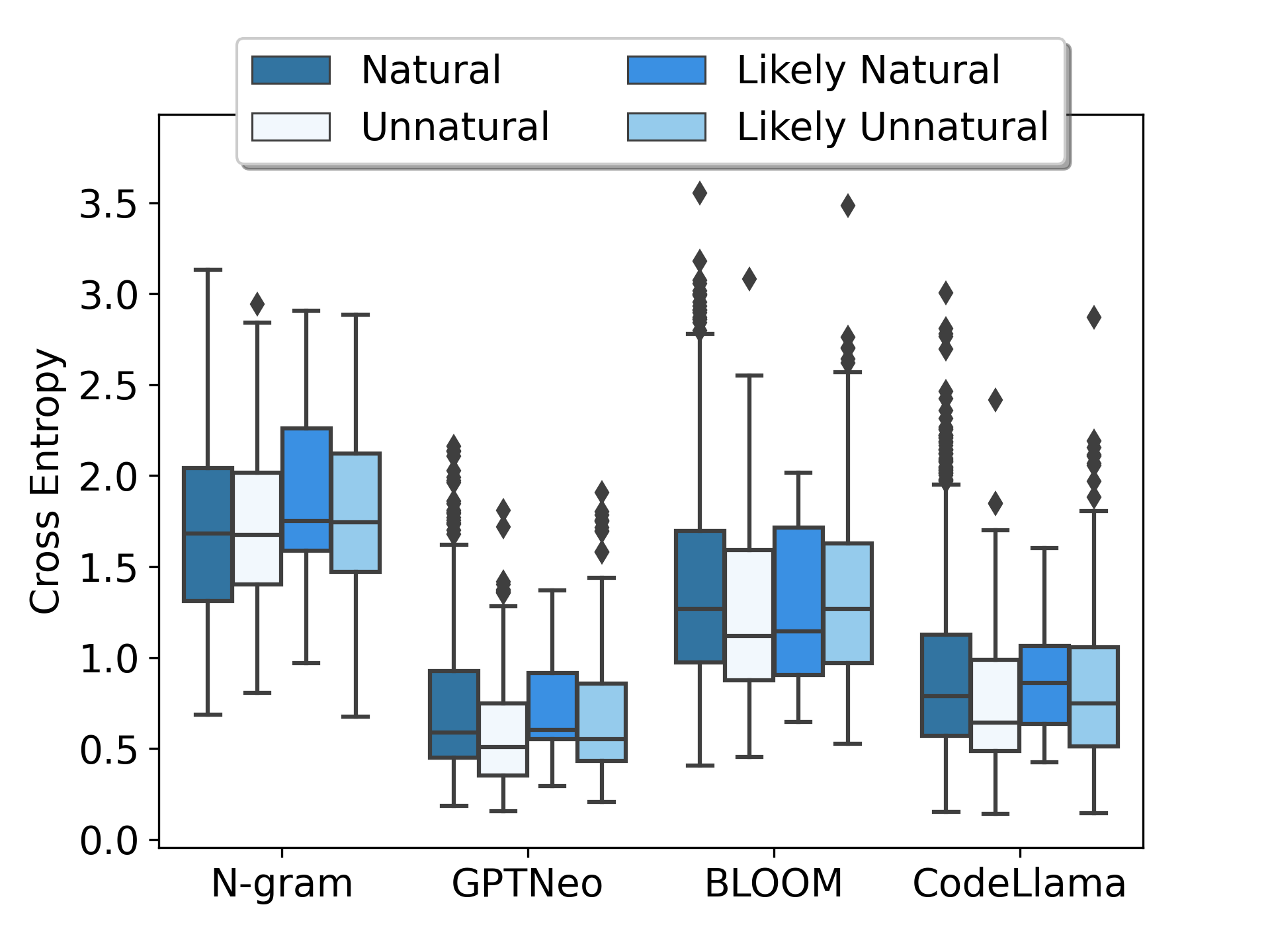}
        \caption{Cross Entropy of Transformed Code}~\label{fig:entropy}
    \end{subfigure}
    \begin{subfigure}[b]{0.48\textwidth}
        \includegraphics[width=\textwidth]{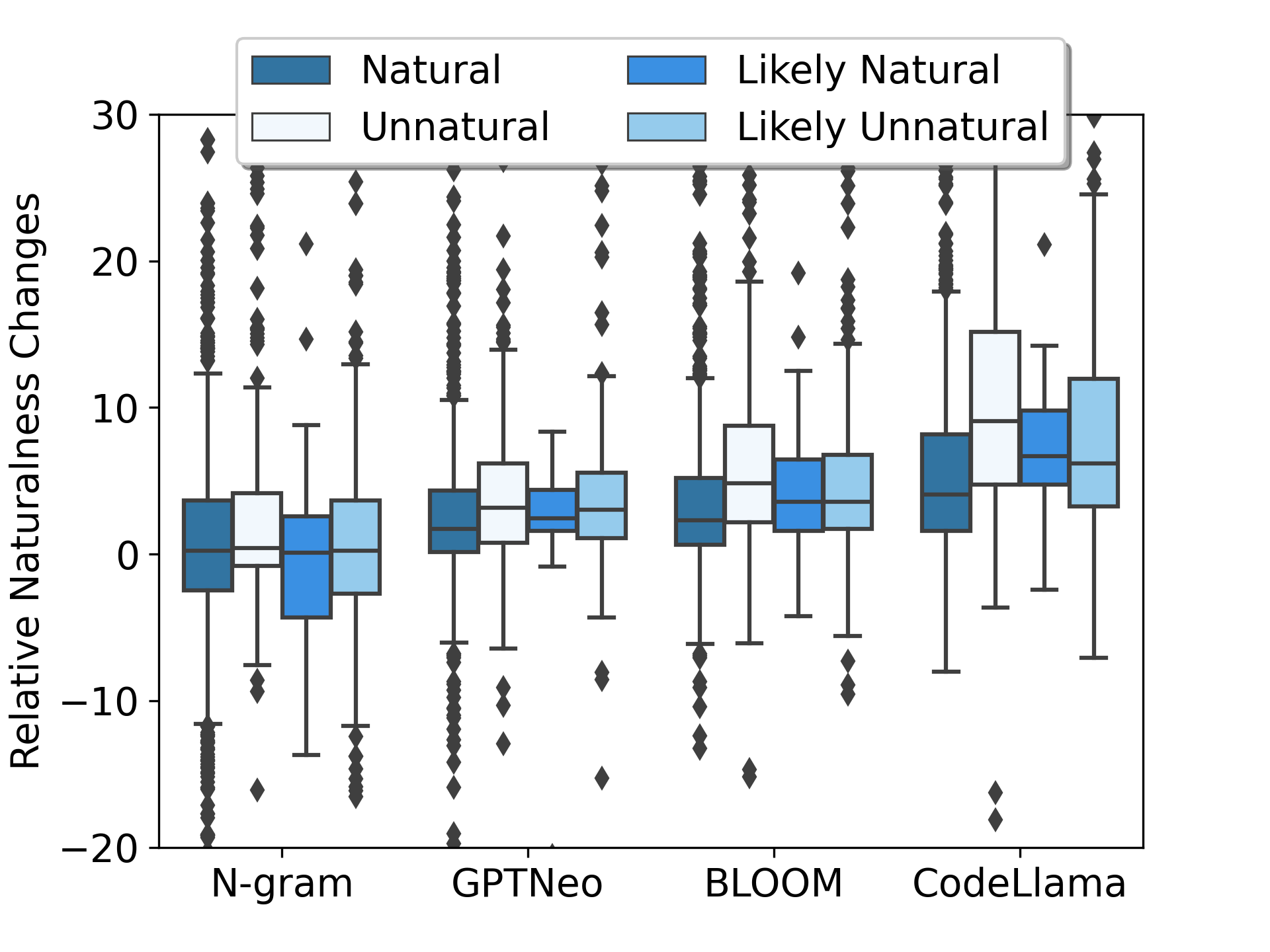}
        \caption{Relative Naturalness Changes}~\label{fig:entropy_compare}
    \end{subfigure}
\end{figure}

In this experiment, we investigate the capability of the existing naturalness metric, specifically Cross-Entropy (CE), calculated by various language models, including N-gram, GPTNeo, BLOOM, and Code Llama, for assessing the naturalness of code transformations. To achieve this goal, we first explore the common approach~\cite{jimenez2018mutants, ray2016naturalness} that directly used the CE values of transformed code as the naturalness of code transformations. Figure~\ref{fig:entropy} illustrates the distributions of CE values for transformed code, categorized into natural and unnatural code transformations in our study.

Overall, our observations reveal that the CE values for transformed code in natural code transformations are unexpectedly higher than those in unnatural ones. This finding contradicts the intuition of the naturalness metric, where lower CE values indicate more natural code. To gain a deeper understanding of this issue, we conducted further analyses and found that the CE values of the transformed code are strongly affected by the CE values of the original code. To validate this observation, we employed Spearman's rank correlation coefficient, a well-known nonparametric measure of rank correlation, to quantify the relationship between CE values of the original and transformed code. Our results yielded a Spearman's correlation coefficient of 0.96 to 0.99 (on different language models), indicating a very strong relationship between the CE values of the original and transformed code. Consequently, the CE value of an unnatural transformed code can even be lower than that of a natural one if the original code has lower CE values. This characteristic makes a naturalness assessment of code transformation, which solely relies on the naturalness of transformed code, yield inaccurate results. 

\find{\textbf{Finding 8:} Relying solely on the naturalness of transformed code to assess code transformations can yield inaccurate results due to the strong correlation between the naturalness of the original and transformed code.}

\subsubsection{Proposed Metric: Relative Naturalness Change}~\label{sec:rnc} Given these insights, we propose another method to assess the naturalness of code transformations, namely \textbf{relative naturalness change (RNC)}, that quantifies the naturalness by calculating the relative differences\footnote{We also considered absolute changes of Cross-Entropy (CE) values between the original code and the code after transformation. This metric, however, is less effective than RNC with an AUC of 0.54-0.63 (compared to 0.55-0.70 produced by RNC). Therefore, in this study, we focus on relative changes.} in Cross-Entropy (CE) values
between the original code and the code after transformation. More formally, let us consider a code transformation denoted as $T$, which takes a software bug $b$ as input and generates a transformed bug $b' = T(b)$. We define the naturalness of the transformation $T$ as 
\begin{equation}
    \text{\textbf{RNC(T)}} = \dfrac{H_{M}(b')-H_{M}(b)}{H_{M}(b)}
\end{equation}
, where $H_{M}$ represents the CE calculated by model $M$. 
This approach allows us to mitigate the impact of the naturalness of the original code on the assessment of code transformations. Figure~\ref{fig:entropy_compare} illustrates the distributions of RNC of unnatural and natural transformations in our study.

From Figure~\ref{fig:entropy_compare}, we can observe that the RNC of unnatural transformations is significantly higher than natural ones on all language models. These observations are also confirmed to be statistically significant by Mann-Whitney-Wilcoxon (MWW) tests with a p-value less than 0.05 with effect size from small (0.09) to medium (0.38). Interestingly, we also note that  RNC also encounters difficulties when dealing with transformations that exhibit human annotator disagreement, i.e., likely to be unnatural or natural transformations. Specifically, the disparity between these two transformation categories is not statistically significant, as indicated by Mann-Whitney-Wilcoxon (MWW) tests yielding p-values ranging from 0.28 to 0.72, suggesting a lack of significant statistical difference. This finding supports our intuition about the sensitivity of naturalness assessment on code transformations in which a subset of these transformations is more difficult to assess by human annotators than others. 

\subsubsection{Effectiveness of Relative Naturalness Change.}~\label{sec:rnc_effectiveness} 
Finally, inspired by the significant differences between the RNC of natural and unnatural transformations, we investigate the capability of using this metric to filter out unnatural code transformations. Particularly, we calculate the probability of code transformation being unnatural based on RNC as 
\begin{equation}
    p_{unnatural} = \frac{RNC- min_{RNC}}{max_{RNC} - min_{RNC}}
\end{equation}

where, $min_{RNC}$ and $max_{RNC}$ is the minimum and maximum values of $RNC$ observed in our dataset. This way normalizes RNC from an arbitrary range into the probability range of  (0, 1). Given this probability, we explore the effectiveness of RNC in identifying unnatural transformations using AUC score, a widely used threshold-independent metric for binary classification tasks. This metric represents the probability that a randomly chosen positive data (i.e., unnatural transformations) will be ranked higher than a randomly
chosen negative one (i.e., natural transformations). As a baseline, we also investigate the performance of Cross-Entropy of transformed code on this task by normalizing this value in the same way with RNC. Particularly,  
\begin{equation}
    p_{unnatural} = \frac{CE- min_{CE}}{max_{CE} - min_{CE}}
\end{equation}
where, $min_{CE}$ and $max_{CE}$ is the minimum and maximum values of $CE$ observed in our dataset.
Table~\ref{tab:rnc_evaluation} shows the effectiveness of CE and RNC with different language models in terms of AUC for filtering out unnatural code transformations. 

\begin{table}[t]
    \centering
    \caption{Evaluation on the effectiveness of Cross-Entropy (CE) and Relative Naturalness Change (RNC) in terms of AUC with different language models for assessing naturalness of code transformations.}
    \label{tab:rnc_evaluation}
    \begin{tabular}{l|cccc}
        \toprule
        \textbf{Methods} & \textbf{Ngram} & \textbf{GPTNeo} & \textbf{BLOOM} & \textbf{CodeLlama}  \\ \midrule
        \textbf{CE} & \textbf{0.49} & 0.40 & 0.42 & 0.40 \\
        \textbf{RNC} & 0.55 & 0.59 & 0.66 & \textbf{0.70}\\ \bottomrule
    \end{tabular}
\end{table}
\vspace{-0mm}
We can see that the combination of RNC with CodeLlama is the best-performing approach with an AUC of 0.7 while the combination of RNC with other models yields an AUC of 0.55-0.66. These results show the promise of using RNC with LLMs such as CodeLlama~\cite{touvron2023llama} for effectively filtering out unnatural code transformations. In comparison to Cross-Entropy, we can see that all combinations of RNC substantially outperform the best combination of CE, i.e., CE + N-gram, in terms of AUC with improvements from 12-43\%. This is because the CE values of the transformed code are significantly affected by the CE values of the original code, making inaccurate naturalness assessment of code transformations. 

\find{\textbf{Finding 9:} Relative Naturalness Changes (RNC) calculated by Language Models yield very promising performance on filtering out unnatural code transformations (with AUC of 0.7)}

\thanh{We investigate the impact of code transformation naturalness and our proposed RNC metric on improving the robustness of NPR techniques. Specifically, we assess how naturalness influences AI model robustness when fine-tuning with data augmentation, a common practice in related work~\cite{zhou2022natural, zhang2022towards}. Although ideally we would test all code transformations, augmenting a program repair dataset using semantic-preserving transformations presents challenges, particularly in aligning ground-truth (fixed) programs with transformations applied to buggy inputs. Addressing this issue is beyond the scope of our study, so we focus our experiments on naming transformations, which are more feasible. Additionally, due to the extensive cost of training models, we conduct our experiment on a single representative NPR model, RepairLLama. We selected RepairLLama for the following reasons: (1) it demonstrates the best performance among NPR techniques, (2) it exhibits the highest prediction changes, and (3) its replication package provides sufficient resources for our experiments, including training scripts and data.}

\thanh{In particular, we enhanced the RepairLLama model’s training dataset with semantic-preserving transformations, fine-tuning it on datasets with varying levels of naturalness, as quantified by RNC. We created six datasets: the original dataset and five augmented sets, corresponding to the top 20\%, 40\%, 60\%, 80\%, and 100\% of transformations based on their RNC scores. To assess the effectiveness of RepairLLama, we used the Exact Match metric, a common evaluation method in previous work, to ensure scalability and efficiency in model development~\cite{chen2019sequencer, silva2023repairllama}.}

\begin{table}[h!]
\centering
\caption{Performance of RepairLLama (measured by Exact Match) fine-tuned on datasets augmented with the top-k\% transformations, ranked by naturalness according to RNC.}
\label{tab:finetune}
\begin{tabular}{c|c|c}
\toprule
\textbf{Top-k} & \textbf{Original Data} & \textbf{Transformed Data} \\ \midrule
0                           & 73.0\%                 & 27.5\%                    \\ 
20                          & 80.3\%                 & 30.3\%                    \\ 
40                          & 72.4\%                 & 27.3\%                    \\ 
60                          & 67.4\%                 & 25.4\%                    \\ 
80                          & 69.1\%                 & 26.1\%                    \\ 
100                         & 61.0\%                 & 23.0\%                    \\ \bottomrule
\end{tabular}
\end{table}

\thanh{Figure~\ref{tab:finetune} illustrates shows the Exact Match performance of RepairLlama fine-tuned on datasets augmented with the top-k\% transformations, ranked by naturalness according to RNC. The best performance is achieved with the top 20\% most natural transformations, where RepairLlama's exact match improves from 73.0\% to 80.3\% on the original data, and from 27.5\% to 30.3\% on the transformed data. However, adding more transformations, especially less natural ones, leads to a notable performance decline on both original and transformed data. We suspect this drop is due to overfitting on recurring bug patterns introduced in the augmented data.}

\thanh{These results suggest that incorporating less natural transformations degrades model performance, likely due to overfitting on repeated bug patterns in the augmented data. In contrast, focusing on the most natural transformations, as ranked by RNC, can significantly boost RepairLlama's performance. While our findings are not fully comprehensive due to the constraints mentioned earlier, we hope they provide a useful starting point for future research.}

\find{\textbf{Finding 10:} Fine-tuning NPR techniques, such as RepairLlama, using top-ranked naming transformations measured by RNC can help the model achieve optimal performance. In constrast, blindly applying less natural transformations degrades model performance, likely due to overfitting on repeated bug patterns in the augmented data.
}
\section{Discussion}
\label{sec:discussion}

\subsection{Implications}
\thanh{
Our study gained insights about natural robustness testing in Neural Program Repair. In this section, we will discuss about implications and suggestions for future works.}

\subsubsection{Natural Robustness Testing for more reliable evaluation of Neural Program Repair} \thanh{Despite the high quality, Program Repair benchmarks often grapple with limited size and infrequent updates, posing challenges in accurately representing real-world bugs. Creating new, large-scale benchmarks is also challenging, as it requires substantial manual effort to maintain quality. Prior work~\cite{gerobustnpr} has proposed using metamorphic (or robustness) testing with semantic-preserving transformations to enhance NPR evaluation reliability without the need for new benchmarks. However, our study reveals that these transformations often involve unnatural modifications that rarely occur in real-world programming, introducing bias into NPR performance assessments and leading to misguided efforts in improving NPR techniques.}

\thanh{To address this, we advocate for natural robustness testing, which evaluates NPR techniques against natural variations in code using natural code transformations. Our experiments reveal unrobust behaviors in NPR techniques when exposed to these variations. More importantly, these unrobustness lead to significant performance decreases and alter empirical findings about NPR performance, particularly regarding improvements of one NPR technique over another. These results underscore the need for more reliable NPR evaluations through natural robustness testing.}

\subsubsection{Naturalness in the Design of Semantic-Preserving Transformations for Natural Robustness Testing}

\thanh{Semantic-preserving transformations play a crucial role in evaluating automated systems for tasks such as AI4Code~\cite{zhou2022natural, zhang2023challenging} and program analysis~\cite{zhang2023statfier, wu2024natural}. These transformations are often used to intentionally trigger edge-case behaviors, helping to uncover errors~\cite{zhang2023statfier, wu2024natural} or expose security vulnerabilities from adversarial attacks~\cite{zhou2022natural, zhang2022towards}. Consequently, many existing techniques prioritize triggering extreme conditions, resulting in transformations that are often unnatural and rarely encountered in real-world programming, leading to a lack of naturalness in the design of semantic-preserving transformations.}

\thanh{In contrast, we argue that naturalness is particularly critical in the context of NPR evaluation, as our goal is to assess natural robustness. This entails evaluating NPR techniques to unveil their true effectiveness against natural variations in code, relying exclusively on natural code transformations. Our study highlights the inherent unnaturalness of many commonly used semantic- preserving transformations, demonstrating their significant influence on both the applicability of these transformations and the conclusions drawn from robustness testing. This underscores the importance of incorporating naturalness in the design of semantic-preserving transformations for effective natural robustness testing.
}

\thanh{
It is worthy to note that while we focus our experiments on the specific domain of Neural Program Repair due to limited resources, our findings on the naturalness of semantic-preserving transformations could potentially generalize across various domains, impacting not only Neural Program Repair but also other tasks. For instance, the use of natural/unnatural code transformations in compiler and analyzer testing~\cite{zhang2023statfier} could help prioritize bug fixing in these systems. Bugs exposed by natural code transformations may be of higher priority, as they are more likely to occur in practice compared to those revealed by unnatural transformations. Therefore, our findings on naturalness assessment criteria and LLM-based naturalness metrics, such as RNC, could be valuable for assisting developers in prioritizing bug resolution efforts.
}

\subsubsection{Automated Naturalness Assessment for Code Transformations.}

\thanh{The naturalness of code transformations often require the extensive manual efforts, hindering their scalability and limiting their scope. In this study, we proposed a novel metric, namely RNC, (see details in Section~\ref{sec:rnc}), for automatically assess naturalness of code transformation using Large Language Models. Our experimental results also demonstrated the effectiveness and potential of this metric. Based on this initial encouraging result, we advocate future research to explore a more comprehensive approach for better-automated naturalness assessment. More specifically, code readability and coding conventions, which are considered highly relevant to naturalness by senior developers in our interviews, could be incorporated into the naturalness assessment to improve its accuracy.}

\subsubsection{Positive impact of Semantic-preserving Transformations}

\begin{lstlisting}[float, language=java, caption=A example of an unnatural bug created by a transformation which divide the infix expression into two sub-expression on the bug Math-63 in Defects4J dataset, escapechar=?, label=lst:positive_example_1]
//Original Code
return (?\hl{Double.isNaN(x) \&\& Double.isNaN(y)}?) || ?\hl{x == y}?;

//Transformed Code
boolean ret = Double.isNaN(x) \&\& Double.isNaN(y);
boolean result = x == y;
return (?\sethlcolor{mGreen}\hl{ret}?) || ?\sethlcolor{mGreen}\hl{result}?;

//Generated Patches for Transformed Code
boolean ret = Double.isNaN(x) && Double.isNaN(y);
boolean result = x == y;
return (!ret) && result;

\end{lstlisting}

\begin{lstlisting}[float, language=java, caption=A example of an unnatural bug created by a variable renaming on the bug JacksonDatabind-76 in Defects4J dataset, escapechar=?, label=lst:positive_example_2]
//Original Code
-  if (buffer.assignParameter(creatorProp, creatorProp.deserialize(p, ?\hl{ctxt}?))) {

//Transformed Code
-  if (buffer.assignParameter(creatorProp, creatorProp.deserialize(p, ?\sethlcolor{mGreen}\hl{context}?))) {

//Generated Patches for Transformed Code
+  if (buffer.assignParameter(creatorProp, creatorProp.deserialize(p, context))) continue;

\end{lstlisting}

\thanh{In the literature, semantic-preserving transformations are frequently employed to trigger errors~\cite{zhang2023statfier, wu2024natural} or expose security vulnerabilities from adversarial attacks~\cite{zhou2022natural, zhang2022towards} in automated systems for code-related tasks. However, in our study, we discovered that these transformations do not always have a negative impact; in some cases, they actually enhance system performance. In the deeper analyses, we suspect that this positive effect may arise because the transformations make the code more aligned with established coding conventions or easier for NPR techniques to repair. For example, in Listing~\ref{lst:positive_example_1},  the transformed code assigns the value of the expression \texttt{Double.isNaN(x) \&\& Double.isNaN(y)} to a variable \texttt{ret}. This transformation simplifies the buggy code, resulting in a reduced search space for AI models. Consequently, RepairLlama can generate patches that pass all test cases by simply negating these values. In contrast, the original code presents challenges for RepairLlama due to the complexity introduced by multiple tokens in the expression \texttt{Double.isNaN(x) \&\& Double.isNaN(y)}. In other example as illustrated by Listing~\ref{lst:positive_example_2}, the variable name \texttt{ctxt} is replaced with a more meaningful name, \texttt{context}, which possibly better aligns with common conventions recognized by Incoder. As a result, this model can correctly fix this error.
This finding motivates further exploration into how code transformations could be strategically used to improve the performance of NPR techniques by inducing positive changes.
}

\subsection{Threats to Validity}
\label{sec:ttv}

\subsubsection{External validity.} These threats concern the
generalizability of our findings. Regarding transformations, we included all transformations generated by 18 operators on 225 bugs from the Defects4J dataset. We believe these transformations are comparable to prior works. Particularly, prior robustness evaluation of AI4Code models only involved from 6 to 17 transformations while existing empirical study~\cite{zhong2022standup4npr} on NPR also contains 260 bugs from Defects4J. Regarding NPR techniques, we originally considered 17 systems but we only can conduct experiments on 5 of them due to availability, configurability, and executability issues. These issues are also observed in the prior works~\cite{zhong2022standup4npr, gerobustnpr}, leading to only 4 and 6 being evaluated. However, we acknowledge that our findings may differ if more tools, transformations, and bugs are considered. 

\vspace{2mm}

\subsubsection{Internal validity.} These threats refer to possible errors in our experiments. To reduce these threats, we carefully check our implementations to avoid potential bugs. Besides, our study involves manual assessments and analyses of (1) interview responses (2) the naturalness of code transformations, and (3) the correctness of patches generated by NPR tools, which may introduce human errors. To minimize the risk, we employ well-known methodologies, including Thematic Analysis~\cite{braun2006using} and Open Card-sorting Discussion~\cite{spencer2009card}, which have been widely used by prior works in software engineering~\cite{liu2023refining, wan2017bug, wan2019does} or human-computer interaction~\cite{gauthier2022computational, cairns2008research, teruel2016applying}. We also involve multiple raters for each task and ensure the quality of assessments and analyses by measuring inter-rater agreement and conducting sanity checks as discussed in Sections ~\ref{sec:criteria} and ~\ref{sec:rq1}. \thanh {Another threat to the internal validity of Neural Program Repair (NPR) in the era of Pre-trained Language Models is the issue of data leakage. Detecting data leakage in NPR techniques is particularly challenging, as the training data for these models is often proprietary. In this work, we consider data leakage as part of the broader robustness challenge. Specifically, since the code transformations we apply are likely absent from the training data, NPR models that rely on memorization may struggle with these transformations, potentially triggering unrobust behaviors. In future work, we could explore seperate these cases by identifying the root causes of robustness failures in examples detected by natural code transformations.}

\vspace{2mm}

\subsubsection{Construct validity.} These threats relate to the suitability of our evaluation. To minimize the risk, we have used well-known evaluation metrics in our study such as Cohen's (Fleiss) Kappa for inter-rater agreement, Mann-Whitney-Wilcoxon for statistical significance tests, Spearman's coefficient for rank correlation, and AUC for classification evaluation. These metrics are commonly used in many research areas, e.g., software engineering~\cite{le2019reliability, jiarpakdee2018autospearman, jiarpakdee2020impact} or information retrieval~\cite{castillo2006reference, amigo2013general}.
\section{Conclusion and Future Work}
\label{sec:conclusion} 

We investigate the impact of naturalness in evaluating Neural Program Repair techniques using semantic-preserving transformations. We conducted a study involving professional developers which resulted in the first concrete criteria for deciding the naturalness of code transformations, which we used in a naturalness assessment of 1178 semantic-preserving transformations. From this study, we found that (1) only 58.8\% of semantic-preserving transformations are deemed natural with high agreement, and, (2) 19.3\% of these transformations are considered unnatural with high agreement. Our experiments on 5 NPR techniques also unveiled that these unnatural transformations introduce a false alarm rate of 25.2\% on the robustness evaluation of these systems. Next, we revisit the robustness of NPR techniques using natural code transformations and found that these systems are still not robust with a drop of from 4.1\% to 14\% and 6.1\% to 23.6\% in terms of the number of correct and plausible patches generated by these systems. Based on these findings we recommend NPR researchers (1) pay more attention to robustness evaluation to study the true effectiveness of NPR techniques on unseen bugs and (2) incorporate Naturalness Assessment for more reliable robustness evaluation. Finally, we proposed Relative Naturalness Changes (RNC), which quantify naturalness by calculating the relative differences in Cross-Entropy (CE) values between the original code and the transformed code. Using RNC, we can automate the assessment of code transformation naturalness, with an AUC of 0.7.

In the future, we plan to expand our research both horizontally and vertically. First, our objective is to investigate the impact of natural robustness of AI4Code models in other Software Engineering tasks, such as vulnerability detection~\cite{li2021vulnerability, chakraborty2021deep}, bug localization~\cite{lou2021boosting, nguyen2022ffl} or program analysis~\cite{10.1145/3597926.3598050, le2022autopruner, le2023invalidator}. \thanh{We also want to extend our research on the impact of unnaturalness of related SE tasks using semantic-preserving transformation such as static analyzer~\cite{zhang2023statfier} and symbolic executor testing~\cite{wu2024natural}.} Second, we plan to broaden the scope of our study by (1) considering additional Program Repair systems, including search-based and semantic-based approaches, and (2) using a wider range of code transformations, considering not only those that affect the buggy line but also those that impact the surrounding buggy context. Finally, given insights from these studies, we plan to investigate a new method to improve the robustness of Program Repair systems.

\bibliographystyle{ACM-Reference-Format}
\bibliography{main}

\newpage


\end{document}